\documentclass[useAMS,usenatbib, usegraphicx]{mn2e}
\usepackage{amssymb}
\usepackage{amsfonts}
\usepackage{graphicx}
\newcommand{\rmd}{{\rm{d}}}
\newcommand{\rme}{{\rm{e}}}

\newcommand{\mH}{m_{\rm{H}}}

\newcommand{\bOmega}{{\mathbf\Omega}}

\newcommand{\beq}{\begin{equation}}
\newcommand{\eeq}{\end{equation}}
\newcommand{\barr}{\begin{eqnarray}}
\newcommand{\earr}{\end{eqnarray}}

\title[Spinning dust]{Spinning dust emission: the effect of rotation around a non-principal axis}

\author[Silsbee, Ali-Ha\"{\i}moud, and Hirata]{
       Kedron Silsbee$^{1}$,
       Yacine Ali-Ha\"{\i}moud$^{2}$\thanks{Corresponding author; e-mail: yacine@tapir.caltech.edu}, and
       Christopher M. Hirata$^{2}$\\
$^{1}$California Institute of Technology, MSC 865, Pasadena, CA 91126, U.S.A. \\
$^{2}$California Institute of Technology, Mail Code 350-17, Pasadena, CA 91125, U.S.A.}

\begin{document}

\date{\today}

\pagerange{\pageref{firstpage}--\pageref{lastpage}} \pubyear{2002}

\maketitle

\label{firstpage}

\begin{abstract}
We investigate the rotational emission from dust grains that rotate around non-principal axes.  We argue that in many phases of the interstellar medium, the smallest 
grains, which dominate spinning dust emission, are likely to have their nutation state (orientation of principal axes relative to the angular momentum vector) 
randomized during each thermal spike.  We recompute the excitation and damping rates associated with rotational emission from the grain permanent dipole, grain-plasma 
interactions, infrared photon emission, and collisions.  The resulting spinning dust spectra generally show a shift toward higher emissivities and peak frequencies
relative to previous calculations.
\end{abstract}

\section{Introduction}

One of the difficulties in measuring the anisotropies in the cosmic microwave background (CMB) is that the interstellar medium (ISM) also emits microwave radiation 
through several mechanisms.  This ``foreground'' radiation must be modeled and subtracted in order to measure the cosmological parameters accurately using the CMB.  The 
standard theory of ISM microwave emission contains three major emission mechanisms \citep[e.g. ][]{2000ApJ...530..133T, 2003ApJS..148...97B, 2009AIPC.1141..265F}: 
synchrotron radiation from relativistic electrons spiralling in the Galactic magnetic field; free-free radiation from ionized gas; and thermal emission from dust 
grains.  These are typically traced by external templates: low-frequency radio maps for the synchrotron \citep{1982A&AS...47....1H}, H$\alpha$ for the free-free 
\citep{2003ApJS..146..407F}, and far-infrared continuum for the dust \citep{1999ApJ...524..867F}.

\citet{Kogut1996a, Kogut1996b} reported a spatial correlation between Galactic microwave emission at 31.5, 53 and 90 GHz and the thermal infrared continuum from dust. 
They interpreted the microwave emission as dust-correlated free-free radiation, on top of the Rayleigh-Jeans tail of thermal emission from dust.
Their observations were confirmed by \citet{deOliveiraCosta1997}, who measured the microwave intensity of the Galaxy at 30 and 40 GHz. \citet{1997ApJ...486L..23L} claimed
the presence of an "anomalous" component of Galactic microwave emission, which they observed as a signal at 14.5 and 32 GHz strongly correlated with the diffuse 100 $\mu$ m intensity. It was far too bright to be thermal dust and had a flat spectrum across these bands, and low-frequency radio and H$\alpha$ observations predicted far too little synchrotron or free-free emission to explain the signal.  
\citet{1997ApJ...486L..23L} proposed that the signal originated from hot gas at $T\ge10^6\,$K, which could produce free-free radiation but little H$\alpha$; however 
\citet{1998ApJ...494L..19D} showed that this gas would cool rapidly and that keeping it hot was energetically unfeasible.  Several alternative explanations have 
been proposed.
Spinning dust emission is due to the rotation of small dust grains with permanent electric dipole moments.  
The basic mechanism has been known for decades \citep{1957ApJ...126..480E,
1970Natur.227..473H, 1992A&A...253..498, 1994ApJ...427..155}, and was suggested as an explanation
for the anomalous emission by \citet{1998ApJ...508..157D} (hereafter DL98b).
Magnetic dust emission is due to thermal fluctuations of the magnetic dipole moments of grains 
including ferromagnetic or ferrimagnetic materials \citep{1999ApJ...512..740D}.
Hard synchrotron radiation would be a new synchrotron component from young (recently-accelerated) high energy electrons, proposed to be strongly correlated 
with the far-infrared emission 
from dust due to their common association with recent star formation \citep{2003ApJS..148...97B}.
Both the spinning and magnetic dust hypotheses predict an emission 
spectrum that peaks in the microwave (the former due to the rotation rates of the smallest grains, and the latter due to the gyrofrequency
in ferromagnetic materials). The hard synchrotron hypothesis is now disfavoured due to the low polarization of the anomalous component observed by the {\slshape 
Wilkinson Microwave Anisotropy Probe} ({\slshape WMAP}; \citealt{2007ApJS..170..335P}), its strong morphological correlation with dust maps 
\citep{2004ApJ...614..186F, 2006MNRAS.370.1125D}, and evidence that the anomalous emission has a rising spectrum at low frequencies ($<20$ GHz; 
\citealt{1999ApJ...527L...9D, 2004ApJ...617..350F, 2005ApJ...624L..89W}).

A key test to distinguish the various models for anomalous emission is to construct predicted emission spectra and compare them to observations.
DL98b computed spinning dust spectra for a variety of interstellar environments, accounting for the main processes that affect grain
rotation: collisions, grain-plasma interactions, infrared emission, and radiation-reaction torque on the grain electric dipole moment.
Model spinning dust spectra have been used extensively to test (and in some cases disfavour or rule out) the spinning dust hypothesis
for the anomalous emission seen in the diffuse high-Galactic latitude ISM \citep{2003ApJS..148...97B, 2004ApJ...614..186F, 2008ApJ...680.1235D, 2009ApJS..180..265G}, 
in the Galactic Plane 
\citep[e.g.][]{2004ApJ...617..350F}, and in dense regions such as molecular clouds
\citep{2004ApJ...614..186F, 2005ApJ...624L..89W, 2006ApJ...639..951C, Casassus2008}
and H{\sc\,ii} regions \citep{2007MNRAS.379..297D, 2009ApJ...690.1585D}, supernova remnants \citep{2007MNRAS.377L..69S},
planetary nebulae \citep{2007MNRAS.382.1607C}, and an external galaxy \citep[NGC6946;][]{2010ApJ...709L.108M}.
\citet{2009ApJ...699.1374D} have even used the anomalous emission seen by {\slshape WMAP} in the warm ionized medium (WIM; traced by H$\alpha$)
to test dust models; they observe a factor of $\sim 3$ {\em lower} anomalous emission than predicted, which they tentatively interpret as
due to depletion of the smallest 
dust grains (the polycyclic aromatic hydrocarbons, or PAHs) in the WIM.

Recently, the grain rotation problem has been revisited by two theoretical groups.  \citet[][hereafter AHD09]{2009MNRAS.395.1055A} constructed a more detailed model 
of
grain rotation, following the angular velocities of grains using a Fokker-Planck equation and re-evaluating the rotational excitation and damping rates
using updated grain properties and a more sophisticated model for the grain-plasma interactions.  They also released a public code, {\sc SpDust}, to
compute spinning dust spectra for any input physical conditions and grain properties.
\citet{2010A&A...509A..12Y} presented a quantum-mechanical
treatment of several of these processes and computed the resulting emission spectra.

The existing theoretical treatments of spinning dust, however, still contain a number of simplifying assumptions.  One of the major uncertainties is the
grain size distribution and typical dipole moment, however this uncertainty can be turned into a virtue by using it to constrain dust models \citep[e.g.][]{2009ApJ...699.1374D}.  Additionally, there are 
uncertainties in the physics of grain rotation, such as the validity of the Fokker-Planck approximation or
the assumed properties such as the evaporation temperature of departing adsorbed atoms.  Some of these pieces of physics are not readily amenable to improvement
by theoretical calculations, but others are.

The purpose of this paper is to revisit the assumption by DL98b and AHD09 that grains rotate around the axis of largest moment of inertia due to internal 
dissipation processes.  We argue in particular that PAHs in the diffuse and high UV flux phases are likely to be in a random nutation state.  This is not a 
trivial detail: a dust grain 
rotating around 
a non-principal axis emits at multiple frequencies, including frequencies well in excess of the instantaneous grain angular velocity.  The fact that electric dipole 
emission depends on the second derivative of the dipole moment $\ddot\bmu$ rather than just $\bmu$ enhances the importance of these higher 
frequencies.\footnote{\citet{2010A&A...509A..12Y} allowed for an arbitrary nutation state, but imposed the assumption that the grain dipole moment be exactly 
parallel to the axis of greatest moment of inertia, which eliminates three of the four frequencies of emission from an axisymmetric grain. They also did not 
re-consider the collisional and plasma excitation and drag coefficients.}
We show in Section~\ref{ss:em} 
that for disc-like grains, at fixed angular momentum incorporating a random nutation state increases the spinning dust emissivity by roughly an order of magnitude.  
Of course, having a random nutation state also modifies the processes that change grain angular momenta.  We investigate each of the major processes and find that the 
typical grain angular momentum is reduced, but still find a factor of 1.6 increase in the peak spinning dust emissivity $j_\nu$ and a factor of 1.3 increase in the 
peak frequency 
for WIM conditions.\footnote{For ease of comparison with previous results, our WIM conditions are those of DL98b: density $n_{\rm H}=0.1\,$cm$^{-3}$, gas 
temperature $T=8000\,$K, 
H ionization fraction $n($H$^+)/n_{\rm H}=0.99$, and radiation field normalization $\chi=1$.}

This paper is organized as follows.  Section~\ref{sec:grain} reviews the key parameters of the grain models.  Section~\ref{sec:rotation} describes the expected 
rotational state of grains and the formalism used in this paper (and in the updated {\sc SpDust}) for describing the grain angular momentum distribution. 
Section~\ref{sec:ed} 
considers the electric dipole emission from grains rotating in a random nutation state.  Subsequent sections consider spin-up and spin-down processes for the grains, 
taking account of nutation: Section~\ref{sec:plasma} considers grain-plasma interactions; Section~\ref{sec:ir} considers infrared photon emission; and 
Section~\ref{sec:coll} considers collisions.  Predicted spinning dust spectra are shown in Section~\ref{sec:results}, where we also explore the sensitivity to some of 
our assumptions.  We conclude in Section~\ref{sec:disc}.

The physical processes affecting grains in non-uniform rotation are very complex, and this paper contains some unavoidably long calculations.  The reader 
interested primarily in the results may skip directly from the end of Section~\ref{ss:em} to the beginning of Section~\ref{sec:results}.

We note that \citet{Hoang10} have recently completed a related analysis in which axisymmetric dust grains are followed through a 2-dimensional space of angular 
velocities $(\omega_\parallel,\omega_\perp)$.  Our analyses agree on the basic conclusion that allowing grains to rotate around a non-principal axis results in an 
increase in the spinning dust emissivity and an increase in the peak frequency.

\section{Grain properties}
\label{sec:grain}

The physical properties of dust grains treated in this paper are unmodified from the model of AHD09.  We briefly summarize the key points here, 
but refer to AHD09 and the references therein for details.

\subsection{Size, shape, and charge}

The grain sizes are described by their volume-equivalent radius $a$, defined by $V=\frac43\pi a^3$.  The fiducial size distribution is taken from 
\citet{2001ApJ...548..296W}.  We consider only the carbonaceous grains because they dominate the population of the smallest grains (typically we find that grains with 
radii $\gtrsim12$\AA\ make no significant contribution to the spinning dust emission).

As in DL98b, the large grains are taken to be spherical and the smallest grains are taken to be planar, as appropriate 
for PAHs, and assume the transition to take place at $a_2 = 6 \ $\AA\ or $N_{\rm C}\approx 100$ carbon atoms.
For simplicity we assume the population of planar grains to be disc-like (although real PAHs can have much more complicated geometries), with a disc radius $R = (\frac43 a^3 d^{-1})^{1/2} 
\approx 7 \textrm{\AA}~ (a/5 \textrm{\AA})^{3/2}$, where we used the interlayer separation in graphite, $d = 3.35\ $\AA, to determine the volume-equivalent radius.
In AHD09, it was found that the treatment of the smallest grains as planar was of only minimal 
importance, resulting in $\sim 10$--20\%\ changes in the emissivity $j_\nu$
near the peak of the spectrum.  This conclusion was however based on the assumption of rotation around 
the axis of greatest moment of inertia \citep[e.g. ][]{1979ApJ...231..404P}, which we argue here is not appropriate.  Indeed, we find a substantial (typically factor 
of $\sim2$) increase in 
the spinning dust emissivity as a consequence of the disc-like geometry of the PAHs.

The grain charge distribution calculation is unmodified from AHD09; it is based on the treatment of charging by electron and ion collisions 
\citep{1987ApJ...320..803D, 2001ApJS..134..263W} and photoelectric charging \citep{2001ApJS..134..263W} assuming a standard interstellar radiation field 
\citep{1982A&A...105..372M, 1983A&A...128..212M} re-scaled by an environment-dependent multiplicative factor $\chi$.

\subsection{Dipole moments}

The grain permanent dipole moment is one of the most uncertain properties as it is not constrained by the UV/optical absorption or IR emission data typically used in 
dust modeling \citep{2001ApJ...548..296W, 2001ApJ...554..778L}.  Our fiducial model is similar to that of AHD09 in assuming a multivariate 
Gaussian distribution (appropriate for the random summation of many bonds with dipole moments) with a root-mean-square value taken from DL98b:
the intrinsic dipole moment is taken to be $\langle\mu_{\rm i}^2\rangle^{1/2}=\beta N_{\rm at}^{1/2}$, where $N_{\rm at}$ is the number of atoms and $\beta$ is a 
normalization factor.  The fiducial value is 0.38$\,$D; this is highly uncertain, although we note that it is reasonable for PAHs that lack exact symmetries, e.g. 
the N-circumcoronene cation sequence (C$_{53}$H$_{18}$N$^+$, a PAH that would have zero dipole moment were it not for the single substitution) has a calculated dipole 
moment corresponding to $\beta=0.16$--1.1$\,$D depending on the position of the substitution \citep{2005ApJ...632..316H}.

For a nonspherical grain it makes sense to consider the orientation of the permanent dipole moment relative to the axis of greatest moment of inertia; that is, we can 
consider both the in-plane dipole moment $\mu_{\rm ip}$ and the out-of-plane moment $\mu_{\rm op}$.  An in-plane dipole moment in a PAH could be produced by e.g. 
nitrogen substitution, as suggested to reproduce the location of the 6.2$\,\mu$m band \citep{2005ApJ...632..316H}, or by incomplete hydrogenation (or 
superhydrogenation) of the peripheral carbon atoms \citep{2003ApJ...584..316L}.  An out-of-plane dipole moment, as assumed by \citet{2010A&A...509A..12Y}, requires 
breaking the mirror-plane symmetry of the PAH, e.g. via warping due to pentagonal rings as occurs in corannulene, C$_{20}$H$_{10}$.\footnote{We note that
searches for corannulene rotational lines in the Red Rectangle have returned null results \citep{2009MNRAS.397.1053P}, but this does not
rule out larger warped PAHs.}

In the absence of a definitive rationale for choosing the dipole moment to be in-plane or out-of-plane, we take for our fiducial model the isotropic ratio $\langle\mu_{\rm 
op}^2\rangle$:$\langle\mu_{\rm ip}^2\rangle=1:2$ (i.e. assign the same moment on all three axes).  This choice is very uncertain, however we find that the resulting spectra are 
only weakly sensitive to it -- e.g. for the fiducial WIM model, we find only a $\sim 12$\%\ change in the characteristic emitted frequency and a $\sim 10$\%\ change
in the total emitted power between the extreme cases of a purely in-plane dipole moment and a purely out-of-plane moment.

\section{Rotation of a disc-like grain}
\label{sec:rotation}

Here we review the formalism to describe the rotation of a general axisymmetric grain, and the physics that determines the nutation angle distribution.

\subsection{General description}

\begin{figure}
\includegraphics[width=3.2in]{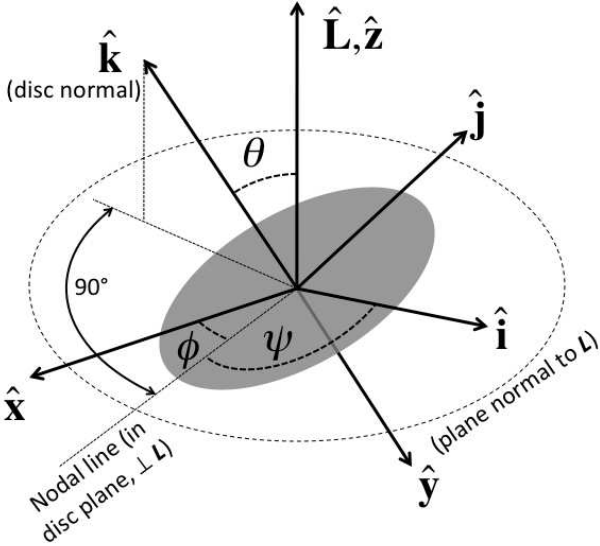}
\caption{\label{fig:euler}The definitions of the Euler angles used in this paper.  
The $xyz$ axes correspond to the inertial frame and the $ijk$ axes to the grain frame.
The angular momentum vector lies in the $\hat{\bmath z}$ direction, the normal to 
the grain disc lies in the $\hat{\bmath k}$ direction, and the grain permanent electric dipole moment lies in the $ik$-plane.}
\end{figure}

We focus here on the case of an oblate axisymmetric dust grain, i.e. one with principal moments of inertia $I_1=I_2<I_3$.  For a planar grain, which is a reasonable 
model for a PAH, one has $I_3=2I_1$.\footnote{For warped PAHs, $I_3/I_1$ is not exactly 2; but it is e.g. 1.93 for corannulene according to the structural 
parameters given in \citet{Hedberg00}.}  In free solid-body rotation, the angular momentum ${\bmath L}$ and rotational energy $E_{\rm rot}$ are conserved; this 
implies 
that the angle $\theta$ between the grain symmetry axis and the angular momentum vector is also conserved.  We may choose the $z$-axis to be along 
the angular momentum vector, so that $\theta$ is one of the Euler angles of the grain (see Fig.~\ref{fig:euler}).  The remaining two Euler angles $\phi, \psi$
then advance at a 
rate (e.g. Eqs. 8.46,47 of \citealt{1998anme.book.....H}):
\begin{equation}
\dot\phi = \frac L{I_1}
\label{eq:dotphi}
\end{equation}
and
\begin{equation}
\dot\psi = -L(I_1^{-1}-I_3^{-1})\cos\theta.
\label{eq:dotpsi}
\end{equation}
We note that $\dot\psi$ may have either sign, but one always has $|\dot\psi|<\dot\phi$.

The rotational energy is given by
\begin{equation}
E_{\rm rot} = \frac{L^2}{2I_1} - \frac{L^2}2(I_1^{-1}-I_3^{-1})\cos^2\theta.
\label{eq:Erot}
\end{equation}

The quantum mechanical description of the system will occasionally be useful (e.g. for counting states) even though the calculation of this paper is in the classical 
regime, as DL98b showed that in general $L \gg \hbar$ even for the smallest grains.\footnote{For the same reason, we neglect issues of nuclear spin statistics
that can arise at small values of $J$ and $K$ for molecules with nontrivial symmetry groups.}
This description is the same as that for a rotating oblate molecule
\citep[e.g.][\S3.9]{Kroto}: the good quantum numbers are the total angular momentum quantum number $J$, with total 
angular momentum $\hbar\sqrt{J(J+1)}$; its projection on the grain 3-axis $\hbar K$; and its projection on the inertial frame $z$-axis $\hbar M$.  These all take 
integer values, with $J\ge 0$ and $|K|,|M|\le J$.  In the case where the lab frame $z$-axis is aligned with the conserved angular momentum, we have $M=J$.  The 
nutation angle $\theta$ satisfies
\beq
\cos\theta = \frac K{\sqrt{J(J+1)}}\approx \frac KJ \ \ \ \ \ \textrm{for $J \gg 1$ .} 
\eeq

\subsection{Rotational configuration}\label{section : rot conf}

The rotational state of PAHs undergoing thermal spikes has been studied in many previous works, particularly those concerned with the polarization of the PAH
emission bands \citep[e.g. ][]{1988prco.book..769L, 2009ApJ...698.1292S}.  Here we recount the key results and explain why we expect PAHs in the diffuse ISM phases to 
generally {\em not} rotate around a 
principal axis of inertia.

\subsubsection{Effect of thermal spikes on grain rotation}

The rotational state of an oblate dust grain is generically described by both an angular momentum $L$ and the angle $\theta$ between this angular momentum and the 
axis of symmetry of the grain.  For large grains, we expect dissipation to bring the grain to the states of minimum rotational energy with at fixed angular momentum, 
i.e. $\theta=0$ or $\theta=\pi$.  For the small grains that dominate spinning dust emission, however, the physics is different because the grain undergoes 
occasional thermal 
spikes (following absorption of each UV photon) followed by cooling into the vibrational ground state.  During thermal spikes, rapid transfer 
of energy is expected to occur between rotational and vibrational degrees of freedom.  This results in a probability distribution for $\theta$:
\begin{equation}
P(\theta|L) d\theta \propto \exp \left[\frac{-E_{\rm rot}(L,\theta)}{kT_{\rm vib}}\right] g(\theta|L) d\theta, 
\end{equation}
where $g(\theta|L)\propto\sin\theta$ is the density of states.\footnote{An easy way to see that the density of states is $\propto\sin\theta$ is to note that at fixed 
total angular momentum $J$, since $K=\sqrt{J(J+1)}\cos\theta$,
the number of states per unit $\cos\theta$ is constant, and hence the number of states per unit $\theta$ is $\propto\sin\theta$.}
This leads to a maximum entropy distribution $P(\theta|L)\propto\sin\theta$ in the limit of $kT_{\rm vib}\gg E_{\rm rot}$, which holds immediately after a UV photon 
absorption.  As the grain cools, $T_{\rm vib}$ drops.  However, as the grain cools, the density of vibrational states drops, and at some temperature $T_{\rm fr}$ the 
vibration-rotation energy transfer freezes out.  We thus expect that the distribution of $\theta$ after a thermal spike freezes out at:
\begin{equation}
P(\theta|L) d\theta \propto \exp \left[\frac{-E_{\rm rot}(L,\theta)}{kT_{\rm fr}}\right] g(\theta|L) d\theta. 
\end{equation}
We consider disc-like grains for $a\le 6\,$\AA\ ($N_{\rm C}\le 100$ carbon atoms).  We note that using the \citet{2001ApJ...551..807D}
model for the vibrational spectrum, the fundamental mode is expected to be at $h\nu_1/k=70(N_{\rm C}/100)^{-1/2}\,$K.  The freeze-out temperature should be at 
least a few times greater than this, depending on the mode spectrum and strength of anharmonic and vibration-rotation couplings.  This is greater than the rotational 
kinetic energy in most of the ISM phases (or similar to it for high radiation density environments such as PDRs).  Thus we expect that in most environments, 
$kT_{\rm fr}$ exceeds the rotational energy, and the
direction of the grain symmetry axis is almost completely isotropized [$P(\theta|L)\propto\sin\theta$] following each thermal spike.



\subsubsection{Frequency of thermal spikes}

Given the major effect of thermal spikes on the rotational state, it is important to consider
how the time between thermal spikes $\tau_{\rm abs}$ compares to the timescale for changes in grain angular momentum $\tau_{\rm rot}$.
The characteristic timescale between UV photon absorptions for a grain
of volume-equivalent radius $a$ is
\beq
\tau_{\rm abs} = \left[\pi a^2 c \int Q_{\rm abs}(a; \nu) \frac{u_{\nu}}{h
  \nu}  \rmd \nu\right]^{-1},
\eeq
where $u_{\nu} = \chi u_{\nu, \rm ISRF}$ is the ambient radiation field and $\pi a^2 Q_{\rm abs}$ is the absorption cross section.

The characteristic rotational damping (or excitation, in steady-state)
timescale for such a grain is $\tau_{\rm rot} \equiv L\big{|}\frac{\rmd L}{\rmd t}\big{|}^{-1}$, where $L$ is the characteristic angular momentum of the grain and $\frac{\rmd L}{\rmd t}$ is the rotational damping rate evaluated at $L$. Evaluating $\tau_{\rm rot}$ requires an analysis of the rotational dynamics. The AHD09 analysis suggests
\beq
\tau_{\rm rot} \approx
\min\left[\frac{\tau_{\rm H}}{F} , \left(\frac{\tau_{\rm H} \tau_{\rm ed}}{G}\right)^{1/2}\right],
\label{eq:taurot}
\eeq
where $F$ and $G$ are the normalized damping and excitation rates; and 
$\tau_{\rm H}$ and $\tau_{\rm ed}$ are the idealized characteristic damping
timescales through collisions with hydrogen atoms and electric dipole
radiation respectively (see AHD09 and the next section for their precise
definitions; and note that $\tau_{\rm ed}$ is defined for thermally rotating grains, but that the actual
dipole damping time varies depending on whether rotation is sub- or super-thermal).\footnote{In Eq.~(\ref{eq:taurot}) the damping time is typically $\tau_{\rm H}/F$ when
linear drag processes dominate.  When electric dipole damping dominates, e.g. in the WIM,
the angular velocity is typically $(G\tau_{\rm ed}/\tau_{\rm H})^{1/4}$ times the thermal angular
velocity $\omega_{\rm th}=(kT/I_3)^{1/2}$ (AHD09).  Since electric dipole emission torque scales as
$\omega^3$ instead of $\omega$, the actual timescale for electric dipole damping is
then $\tau_{\rm ed}(\omega/\omega_{\rm th})^{-2}$, or $(\tau_{\rm H}\tau_{\rm ed}/G)^{1/2}$.}

Since the smallest grains rotate fastest and determine the peak of the spinning dust
spectrum, we evaluate the above timescales at the smallest grain size
$a = 3.5\,$\AA, for the idealized interstellar environments defined in
DL98b, Table 1. We show these timescales in Table \ref{tab:timescales},
for both the case of $\theta=0$ (AHD09) and for isotropized $\theta$ (using the formulae in this paper).
\begin{table*}
\caption{
        \label{tab:timescales}
        Characteristic timescales for UV photons absorption and
        rotational damping for idealized interstellar phases. The
        rotational damping time is shown for grains rotating
        about their axis of greatest inertia (``case 1'', as assumed
        in DL98b, AHD09), and for grains which are randomly oriented
        with respect to their angular momentum (``case 2'', the subject of the
        present work).  All values are for the smallest grains ($a=3.5\,$\AA~or $N_{\rm C}=20$).
        }
\begin{tabular}{cccccccc}
\hline\hline
Phase                   &DC     &MC     &CNM    &WNM    &WIM  &RN   &PDR        \\
\hline
$\tau_{\rm abs}$ (sec)  &$2.0\times10^{11}$     &$2.0\times10^9$        &$2.0\times10^7$        &$2.0\times10^7$        &$2.0\times10^7$  &$2.0\times10^4$  &$6.6\times10^3$    \\
$\tau_{\rm rot}$        (sec) [case 1]  &$1.6\times10^7$        &$9.5\times10^7$        &$1.9\times10^8$        &$2.8\times10^8$
&$2.1\times10^8$ &$7.0\times10^6$  &$1.4\times10^6$     \\
$\tau_{\rm rot}$ (sec) [case 2] &$1.4\times10^7$        &$4.1\times10^7$        &$8.2\times10^7$        &$1.2\times10^8$
&$9.0\times10^7$ &$6.9\times10^6$  &$1.1\times10^6$     \\
\hline\hline
\end{tabular}
\end{table*}

In the diffuse ISM phases (CNM, WNM, WIM), thermal spikes occur with a
rate at least $\sim 4$ to 6 times higher than the processes that change the grain angular
momentum. The rate difference is even more pronouced in regions
of high radiation intensity (RN, PDR), where the small grains can
absorb several hundreds of photons during the time it takes to change
their angular momentum. Therefore we expect an
isotropic distribution $P(\theta|L) \propto \sin \theta$ in these
phases. Note that this is \emph{not}
true of regions of lower radiation density (DC, MC), where thermal
spikes occur every few hundreds to thousands of years and $\tau_{\rm
  abs} \gg \tau_{\rm rot}$. In such cases, other processes will dominate the distribution
of $\theta$ and the result may be in between complete isotropization (as assumed here) and perfect rotation around the $I_3$
axis ($\theta = 0$; assumed in DL98b and AHD09).
An example of such an intermediate case would be the Maxwellian 
distribution for $\theta$ \citep{1967ApJ...147..943J, 1997ApJ...484..230L}.

\subsection{Angular momentum distribution}

The previous spinning dust analysis by AHD09 followed the Fokker-Planck equation for the probability distribution of grains as a 
function of their angular velocity vector $\bomega$.  Since $\bomega$ is not conserved for a nonspherical grain, the proper variable to follow instead 
is the angular 
momentum ${\bmath L}$.  However, in order to maintain a simple connection to previous work, we define the variable:
\begin{equation}
\bOmega \equiv \frac{\bmath L}{I_3}.
\end{equation}
This is the angular velocity that the grain would have {\em if} it were able to dissipate the energy associated with its nutation; we note that the 
magnitude of the actual angular velocity $\bomega$ satisfies $|\bomega|\ge|\bOmega|$.  In this 
paper, the Fokker-Planck equation is constructed in terms of $\bOmega$.

For disc-like grains considered in this paper, with $I_1=\frac12I_3$, the rotational rates become:
\beq
\dot\phi = 2\Omega
{\rm ~~and~~}
\dot\psi = -\Omega\cos\theta.
\eeq
These results will be needed repeatedly throughout the paper.

\subsubsection{Form of the Fokker-Planck equation}

Following the treatment of AHD09, we write the general Fokker-Planck equation for the equilibrium distribution of $\bOmega$:
\beq
\frac{\partial}{\partial \Omega^i}\left[D^i(\bOmega) f_a(\bOmega)
\right]  + \frac{1}{2} \frac{\partial^2}{\partial \Omega^i \partial \Omega^j}
\left[ E^{ij}(\bOmega) f_a(\bOmega) \right]
 = 0.
\eeq 
The Fokker-Planck coefficients are
\beq
D^i(\bOmega) \equiv - \lim_{\delta t \rightarrow 0} \frac{\langle\delta  \Omega^i \rangle}{\delta t}
{\rm ~~and~~} E^{ij}(\bOmega) \equiv \lim_{\delta t \rightarrow 0} \frac{\langle \delta \Omega^i \delta \Omega^j \rangle}{\delta t}.
\label{eq:DEcoefs}
\eeq
Here ${\bmath D}$ denotes the mean drift in $\bOmega$, and ${\mathbfss E}$ denotes the diffusion coefficient tensor.

It is important to note that, because of the isotropic distribution of the direction of the grain symmetry axis (see Section~\ref{section : rot conf}), these 
coefficients are averaged over the angle $\theta$. More explicitly, 
\beq
D^i(\bOmega) \equiv - \frac{1}{2}\int_{0}^{\pi} \lim_{\delta t \rightarrow 0} \frac{\langle\delta  \Omega^i \rangle}{\delta t}\big{(}\bOmega, \theta\big{)}\sin \theta \rmd \theta \ ,
\eeq
and similarly for $E^{ij}(\bOmega)$.

We now assume an isotropic
medium, which is a good approximation so long as we are considering the total intensity spectrum (small deviations from isotropy would result in
net polarization, which is not the subject of this paper).  The drift and diffusion terms can then be decomposed as
\beq
{\bmath D}(\bOmega) = D(\Omega)\hat{\bmath e}_{\bOmega}
\eeq
and
\beq
{\mathbfss E}(\bOmega) = E_{\parallel}(\Omega) \hat{\bmath e}_{\bOmega}\otimes \hat{\bmath e}_{\bOmega}
  + E_\perp(\Omega) ({\mathbfss 1}-\hat{\bmath e}_{\bOmega}\otimes \hat{\bmath e}_{\bOmega}),
\eeq
where $\hat{\bmath e}_{\bOmega}$ is the unit vector in the direction of $\bOmega$ and ${\mathbfss 1}$ is the identity matrix.  The function $D(\Omega)$ then denotes
the rate of damping of rotation, while $E_\parallel(\Omega)$ and $E_\perp(\Omega)$ measure random excitation of the magnitude and direction of the angular momentum
vector.  AHD09 then show that the overall distribution function for $\Omega$ satisfies the equation
\beq
\frac{\rmd f_a(\Omega)}{\rmd \Omega} + 2 \frac{\tilde{D}(\Omega)}{E_{\parallel}(\Omega)} f_a(\Omega) = 0,
\label{eq:FP}
\eeq
where
\beq
\tilde D(\Omega) \equiv D(\Omega) + \frac{E_{\parallel}(\Omega) - E_{\perp}(\Omega)}\Omega + \frac{1}{2}\frac{\rmd E_{\parallel}(\Omega)}{\rmd \Omega}.
\label{eq:Dtilde}
\eeq
Note that $\tilde D$ is simply equal to $D$ if the excitation rates are isotropic and independent of $\Omega$.  This is true for some of the mechanisms described,
but plasma excitation in particular has nontrivial $\Omega$ dependence and here Eq.~(\ref{eq:Dtilde}) is necessary.

\subsubsection{Excitation and damping coefficients}

The $\tilde D(\Omega)$ and $E_\parallel(\Omega)$ are sufficient to write the Fokker-Planck equation but are nontrivial to interpret and vary wildly as a function of
grain size.  For this reason, DL98b introduced dimensionless coefficients $F$ and $G$ that describe damping and excitation rates relative to those
that one would obtain from the ballistic impact of hydrogen atoms on an idealized spherical grain.
These are, for process $X$,
\beq
F_X(\Omega) \equiv \frac{\tau_{\rm H}}{\Omega} \tilde D_X(\Omega)
\label{eq:F}
\eeq 
and
\beq 
G_X(\Omega) \equiv \frac{I_3 \tau_{\rm H}}{2 k T} E_{\parallel,X}(\Omega),
\label{eq:G}
\eeq
where $\tau_{\rm H}$ is the idealized damping timescale (whose precise definition is given in AHD09) and $T$ is the gas temperature.

\subsubsection{Fluctuation-dissipation theorem}

In their analysis of spherical grains, DL98b and AHD09 argued that processes resulting from interaction with a thermal 
bath at temperature $T_X$ (notably plasma drag and excitation) should obey the fluctuation-dissipation theorem, $\tilde D = I_3\Omega E_{\parallel}/2kT_X$.
The equivalent 
result for excitation and damping coefficients is that $F=(T/T_X)G$. No such result can apply here because the randomization of the nutation degree of freedom during 
thermal spikes renders the notion of a ``thermal'' distribution for $\Omega$ not internally consistent.  However, the fluctuation-dissipation theorem's close cousin, 
the principle of detailed balance, can be of some use if one computes damping and excitation of the actions $|{\bmath L}|\approx \hbar J$ and $L\cos\theta=\hbar K$ 
for individual $(J,K)$ levels, and then averages the resulting coefficients over $K$.  We will need to use this technique to compute the plasma drag on a grain 
rotating around a non-principal axis.

\section{Electric dipole emission}
\label{sec:ed}

Our task in computing the emission spectrum falls into two major steps.  One is to relate the distribution of rotational states $f_a(\Omega)$ to the observable 
emission.  The other, harder task, is to compute the Fokker-Planck coefficients arising from each mechanism.  We consider the emission process in this section, and 
then proceed to consider the damping and excitation mechanisms in later sections.

For the case of a grain rotating around a principal axis of inertia, the grain merely rotates with constant angular velocity $\bomega$ and emits monochromatic radiation at frequency $\omega/(2\pi)$.  
Thus in these models (DL98b, AHD09) the emitted spectrum from a particular grain is built up from its dipole moment and the probability distribution for $\omega$.  The 
non-uniform rotation case treated here is more complicated, as we will see that four frequencies are emitted.

\subsection{Emission spectrum}
\label{ss:em}

Our first step in the analysis is to consider how the electric dipole moment $\bmu$ of a grain varies as a function of time.  We define the $\hat{\bmath i}$, $\hat{\bmath j}$, and $\hat{\bmath k}$ 
vectors to form a grain-fixed basis with $\hat{\bmath k}$ along the symmetry axis.  Without loss of generality, $\bmu$ may be assumed to be in the plane defined by $\hat{\bmath i}$ and $\hat{\bmath 
k}$.  Then:
\beq
\bmu = \mu_{\rm ip}\hat{\bmath i} + \mu_{\rm op}\hat{\bmath k},
\label{eq:mu}
\eeq
where $\mu_{\rm ip}$ and $\mu_{\rm op}$ are the in-plane and out-of-plane components of the dipole moment, respectively.

We now consider the behavior of the dipole moment relative to an inertial coordinate system.  We choose the inertial $\hat{\bmath z}$ axis to be parallel to the angular momentum; then we define the 
$3\times 3$ orthogonal matrix ${\mathbfss U}$ with elements $U_{xi}=\hat{\bmath x}\cdot\hat{\bmath i}$, and similarly for the other 8 entries.  The entries involving $\hat{\bmath i}$ and $\hat{\bmath
k}$ are needed here:
\barr
U_{xi} &=& \cos{\phi}\cos{\psi} - \cos{\theta}\sin{\psi}\sin{\phi},
\nonumber \\
U_{yi} &=& \cos{\psi}\sin{\phi} + \cos{\phi}\cos{\theta}\sin{\psi},
\nonumber \\
U_{zi} &=& \sin{\psi}\sin{\theta},
\nonumber \\
U_{xk} &=& \sin{\theta}\sin{\phi},
\nonumber \\
U_{yk} &=& - \sin{\theta} \cos{\phi}, {\rm ~~and}
\nonumber \\
U_{zk} &=& \cos{\theta}.
\label{eq:U}
\earr
For our purposes, it is most convenient to express the first two of these using the product-to-sum rule:
\barr
U_{xi} &=& \frac12[(1-\cos\theta)\cos{(\psi - \phi)}
\nonumber \\ &&
 + (1+\cos{\theta})\cos(\psi+\phi)] {\rm ~~and}
\nonumber \\
U_{yi} &=& \frac12[(1+\cos\theta)\sin{(\psi+\phi)}
\nonumber \\ &&
 + (1-\cos\theta)\sin(\phi-\psi)].
 \label {eq:UPTS}
\earr
The advantage of this formulation is that since $\dot\psi$ and $\dot\phi$ are constant, we have expressed all required components of ${\mathbfss U}$ as sinusoidal functions of time.  Each sinusoidal 
function directly emits a $\delta$-function spectrum at its frequency.  One can see that the above components of ${\mathbfss U}$ oscillate with the four (angular) frequencies $\dot\phi$, $|\dot\psi|$, 
$\dot\phi+\dot\psi$, and $\dot\phi-\dot\psi$.  From Eq.~(\ref{eq:mu}) we see that the same frequencies are present in $\bmu$ (as observed in inertial coordinates).

The power emitted by an accelerating dipole is given by
\beq
P = \frac{2\ddot\bmu^2}{3c^3}
\eeq 

From Eqs. $(\ref{eq:mu})$ and $(\ref{eq:UPTS})$, we see we may write $\ddot\bmu$ as
\barr
\ddot\bmu \!\! &=& \!\! \Bigl\{ -\frac12\mu_{\rm ip}[(1-\cos{\theta}) (\dot\psi - \dot\phi)^2\cos(\psi-\phi)
\nonumber \\ && \!\!
+(1+ \cos{\theta})(\dot\psi + \dot\phi)^2
\cos(\psi + \phi)]
\nonumber \\ && \!\!
 - \mu_{\rm op}\dot\phi^2 \sin{\theta}\sin{\phi} \Bigr\} \hat{\bmath x} 
\nonumber \\ && \!\!
+\Bigl\{ -\frac12\mu_{\rm ip} [(1+ \cos{\theta}) (\dot\psi + \dot\phi)^2 \sin{(\psi + \phi)}
\nonumber \\&& \!\!
+ (1-\cos{\theta})(\dot\phi - \dot\psi)^2\sin{(\phi - \psi)}
]
\nonumber \\ && \!\!
+ \mu_{\rm op}\dot\phi^2\sin{\theta}\cos{\phi} \Bigr\} \hat{\bmath y}
\nonumber \\ && \!\!
- \mu_{\rm ip}\dot\psi^2 \sin{\theta}\sin{\psi}\, \hat{\bmath z}.
\label{eq:muderiv}
\earr
We observe that when we average over many cycles of $\phi$ and $\psi$, all terms average to zero except those which can be expressed in terms of just $\sin^2{\omega}$  or $\cos^2{\omega}$, where $\omega$ is one of $\dot\phi$, $\dot\psi$, $\dot\phi - \dot\psi$ of $\dot\phi + \dot\psi$.  Each of these terms contributes power which is emitted at frequency $\omega$.  We find that
\begin{itemize}
\item
At frequency $\dot\phi + \dot\psi$, the emitted power is
\beq
P_{\dot\phi + \dot\psi} = \frac{\mu_{\rm ip}^2 (\dot\psi + \dot\phi)^4 (1+\cos{\theta})^2}{6c^3}.
\label{eq:comp1}
\eeq
\item
At frequency $\dot\phi - \dot\psi$, the emitted power is
\beq
P_{\dot\phi - \dot\psi} = \frac{\mu_{\rm ip}^2 (\dot\phi - \dot\psi)^4 (1-\cos{\theta})^2}{6c^3}.
\eeq
\item
At frequency $\dot\phi$, the emitted power is
\beq
P_{\dot\phi} = \frac{2\mu_{\rm op}^2 \dot\phi^4 \sin^2\theta}{3c^3}. 
\eeq
\item
At frequency $|\dot\psi|$, the emitted power is
\beq
P_{|\dot\psi|} = \frac{\mu_{\rm ip}^2 \dot\psi^4 \sin^2\theta}{3c^3}.
\eeq
\end{itemize}

The overall emitted spectrum from a grain of given angular momentum $L$ is then obtained by finding the amount of power emitted in a range of angular frequencies $(\omega,\omega+\rmd\omega)$ using both the emitted power for each of the 4 components and the probability of that component falling in the range $(\omega,\omega+\rmd\omega)$.  Consider for example the $\dot\phi + \dot\psi$ component.  
Letting $\omega = \dot\phi + \dot\psi$, we can see that $\omega$ is bounded by:
\beq
\frac{L}{I_3}\le \omega \le 2\frac{L}{I_1} - \frac{L}{I_3} , \ \ \ \textrm{i.e.} \ \ \ \Omega \le \omega \le 3 \Omega. 
\label{eq:bound}
\eeq
Within this range, the probability distribution for $\omega$ can be found using
\beq
\omega = \dot\phi + \dot\psi = \frac L{I_1} - \left(\frac L{I_1}-\frac L{I_3}\right)\cos\theta = \Omega(2 - \cos \theta). 
\eeq
Since $\cos\theta$ is uniformly distributed between $-1$ and $1$ with density $\frac12$, the probability density for $\omega$ is then
\beq
{\rm Prob}(\omega)\rmd\omega = \frac12\left(\frac L{I_1}-\frac L{I_3}\right)^{-1} \rmd\omega = \frac{1}{2} \frac{\rmd \omega}{\Omega} 
\eeq
and the nutation angle that corresponds to emission at $\omega$ is
\beq
\theta = \arccos \frac{L/I_1 - \omega}{L/I_1-L/I_3} = \arccos\left(2 - \frac{\omega}{\Omega}\right).
\eeq
The overall emission spectrum for the $\dot\psi + \dot\phi$ component is then Prob$(\omega)$ times the power at this component, Eq.~(\ref{eq:comp1}); this is\footnote{Note that $P(\omega)$ has units of 
ergs per second per (radian per second) per grain.}
\barr
P_{\dot\phi + \dot\psi}(\omega) &=& \frac{\mu_{\rm ip}^2 \omega^4[1+(L/I_1 - \omega)/(L/I_1-L/I_3)]^2}
{12c^3(L/I_1-L/I_3)}\\
&=& \frac{\mu_{\rm ip}^2 \omega^4 \left( 3 - \omega/\Omega\right)^2}{12 c^3 \Omega}. 
\label{eq:p1}
\earr
A similar calculation shows that we obtain the same spectrum from emission at $\dot\phi - \dot\psi$; this is to be expected since the two components are related by the symmetry 
$\theta\leftrightarrow\pi-\theta$.  Thus:
\beq
P_{\dot\phi - \dot\psi}(\omega) = P_{\dot\phi + \dot\psi}(\omega). 
\label{eq:p2}
\eeq
Following the same procedure, we find that the spectrum emitted at $|\dot\psi|$ is given by
\barr
P_{|\dot\psi|}(\omega) &=& \frac{\mu_{\rm ip}^2 \omega^4 [ 1 - \omega^2/(L/I_1-L/I_3)^2 ]}{3c^3(L/I_1-L/I_3)} \\
&=& \frac{\mu_{\rm ip}^2 \omega^4 \left( 1 - \omega^2/\Omega^2 \right)}{3c^3\Omega} 
\label{eq:p3} 
\earr
within the range $0\leq\omega\leq L/I_1-L/I_3$, i.e. $0 \le \omega \le \Omega$.

Finally the $\dot\phi$ component is at angular frequency $L/I_1 = 2 \Omega$, irrespective of $\theta$.  As calculated before, the total power emitted at this frequency is $4\mu_{\rm op}^2(L/I_1)^4/(9c^3)$.  Thus 
the emitted spectrum is
\beq
P_{\dot\phi}(\omega) = \frac{4\mu_{\rm op}^2\omega^4}{9c^3}\delta\left(\omega-\frac L{I_1}\right) = \frac{4\mu_{\rm op}^2\omega^4}{9c^3}\delta\left(\omega-2 \Omega\right). 
\label{eq:p4}
\eeq

The total emitted spectrum is then the sum of the 4 components, Eqs.~(\ref{eq:p1}--\ref{eq:p4}), considered only within their respective range of validity.  In the particular case of $I_1=\frac12I_3$, we 
see that $L/I_3=\Omega$, $L/I_1=2\Omega$, and
\barr
P(\omega|\Omega) &=& \frac{\omega^4}{c^3}\Bigl\{ \frac{\mu_{\rm ip}^2}{6\Omega}\left( 3 - \frac\omega\Omega\right)^2  
\chi_{\Omega<\omega<3\Omega}
\nonumber \\ && 
+
 \frac{\mu_{\rm ip}^2}{3\Omega} \left( 1-\frac{\omega^2}{\Omega^2}\right)\chi_{\omega<\Omega}
\nonumber \\ &&
+
\frac49\mu_{\rm op}^2\delta(\omega-2\Omega)
\Bigr\},
\label{eq:emSpec}
\earr
where the truth function $\chi$ is 1 if the subscripted inequality holds and 0 otherwise.
The total power emitted per grain is then
\beq
\dot E_{\rm spdust} = \frac{2\Omega^4}{3c^3} \left( 5\mu_{\rm ip}^2 + \frac{32}3\mu_{\rm op}^2 \right).
\label{eq:totem}
\eeq
This should be compared to $2\Omega^4\mu_{\rm ip}^2/(3c^3)$ for the case of a grain rotating around the $\hat{\bmath k}$-axis; for an in-plane dipole moment 
($\mu_{\rm op}=0$) the emitted power is $5$ times higher, whereas for an isotropically distributed dipole moment ($\mu_{\rm ip}^2:\mu_{\rm op}^2=2:1$)
the emitted power is $\sim 10$ times higher.

The emissivity per H atom $j_{\nu}$ (units of erg$\,$s$^{-1}\,$Hz$^{-1}\,$sr$^{-1}$ per H atom) can then be obtained by integrating over the probability distribution for $\Omega$ and the grain size 
distribution:
\barr
j_{\nu} &=& \frac{1}{2} \int \rmd a \frac{1}{n_{\rm H}}\frac{\rmd n_{\rm gr}}{\rmd a}\int \rmd \Omega   P(\omega | \Omega)  4 \pi \Omega^2 f_a(\Omega) 
\nonumber\\
&=& \frac{1}{2} \frac{\omega^4}{c^3} \int \rmd a \frac{1}{n_{\rm H}}\frac{\rmd n_{\rm gr}}{\rmd a} 
\nonumber\\
&&\times \Bigl\{\frac{\mu_{\rm ip}^2}{6}\int_{\frac\omega 3}^{\omega} \frac{\rmd \Omega}{\Omega} \left( 3 - \frac\omega\Omega\right)^2  4 \pi \Omega^2  f_a(\Omega) 
\nonumber\\
&&+ \frac{\mu_{\rm ip}^2}{3} \int_{\omega }^{\infty} \frac{\rmd \Omega}{\Omega} \left( 1-\frac{\omega^2}{\Omega^2}\right)  4 \pi \Omega^2 f_a(\Omega)  
\nonumber\\
&&+ \frac{2 \mu_{\rm op}^2}{9} \pi \omega^2 f_a\left(\frac{\omega}{2}\right)\Bigr\} , 
\earr
where $\omega = 2 \pi \nu$ and the factor $1/2$ comes from multiplying by $2 \pi$ (conversion from $\omega$ to $\nu$) and dividing by $4\pi$ (per steradian).

\subsection{Radiation-reaction torque}

We also need the torque $-{\bmath T}_{\rm rad}$ radiated by the tumbling dipole.  This radiation back-reacts on the grain, applying a radiation-reaction torque $+{\bmath T}_{\rm rad}$.
The general formula for this torque is
\beq 
{\bmath T}_{\rm rad} =  -\frac{2}{3c^3} \langle \dot\bmu \times\ddot\bmu\rangle,
\eeq
where $\langle...\rangle$ denotes a time average.  Since the rotation of a rigid solid body is quasiperiodic, this amounts to first an average over $\phi$ and $\psi$; and in our case, also an average 
over $\cos\theta$ because of the rapid redistribution of the nutation angle.  Averaging over $\phi$ immediately implies that the $x$ and $y$ components of ${\bmath T}_{\rm rad}$ vanish; the $z$-component 
is, after extensive but straightforward manipulation of trigonometric functions,
\barr
T_{{\rm rad},z} &=& \frac{L^3}{24c^3I_1^3I_3^3} \Bigl\{
-3(2I_1^3 + I_1^2I_3 +I_3^3)\mu_{\rm ip}^2 - 8I_3^3\mu_{\rm op}^2 
 \nonumber\\ &&
+ [-8I_1^3\mu_{\rm ip}^2
+4I_3^3(\mu_{\rm ip}^2 + 2\mu_{\rm op}^2)]\cos{2\theta}
 \nonumber\\ &&
 - (I_1 - I_3)^2(2I_1 + I_3)\mu_{\rm ip}^2\cos{4 \theta}
\Bigr\}.
\earr
Averaging over nutation angles (by multiplying by $\frac12\sin\theta$ and integrating over $0<\theta<\pi$) gives
\beq
T_{{\rm rad},z} = \frac{-2L^3(3I_1^3 + 3I_1^2I_3 + 4I_3^3)\mu_{\rm ip}^2 + 20L^3I_3^3\mu_{\rm op}^2} {45 I_1^3 I_3^3 c^3}.
\eeq
The case of interest here is $I_1=\frac12I_3$, for which
\beq
T_{{\rm rad},z} = -\frac{\Omega^3}{c^3}\left(
\frac{82}{45}\mu_{\rm ip}^2 +  \frac{32}{9}\mu_{\rm op}^2
\right).
\label{eq:Torque}
\eeq
This compares with $-2\Omega^3\mu_{\rm ip}^2/(3c^3)$ for the uniformly rotating case.

We note that Eq.~(\ref{eq:Torque}) can also be obtained semiclasically by noting that the photons emitted in the $\dot\phi$ and $\dot\phi\pm\dot\psi$ frequencies carry $z$ angular momentum of $+\hbar$ 
per photon, 
while those emitted at the $|\dot\psi|$ frequency carry no $z$ angular momentum.\footnote{This can be seen by observing that the dipole components at frequencies 
$\dot\phi$ and $\dot\phi\pm\dot\psi$ are rotating in the $xy$-plane, while that at $|\dot\psi|$ is oscillating along the $z$-axis.}  The ratio of 
angular momentum radiated to energy radiated is thus $\omega^{-1}$ for the $\dot\phi$ and $\dot\phi\pm\dot\psi$ components, and we could have written
\beq
T_{{\rm rad},z} = -\int \omega^{-1} P_{\dot\phi,\dot\phi\pm\dot\psi}(\omega) \rmd\omega.
\eeq
This argument, combined with Eq.~(\ref{eq:emSpec}), confirms Eq.~(\ref{eq:Torque}).

Radiation-reaction is implemented in {\sc SpDust} using the electric dipole damping time $\tau_{\rm ed}$, defined by DL98b to be the 
radiation-reaction damping 
time $L/(2|T_{{\rm rad},z}|)$ 
for a grain rotating at thermal velocity, i.e. with rotational kinetic energy $\frac32kT$, about the axis of greatest inertia
.  Mathematically:
\beq
\left.\frac{\rmd\Omega}{\rmd t}\right|_{\rm rad-reac} = -\frac{I_3\Omega^3}{3kT\tau_{\rm ed}}.
\eeq
Our calculation establishes that the damping time for planar axisymmetric grains is
\beq
\tau_{\rm ed} = \frac{I_3^2c^3}{3kT} \left( \frac{82}{45}\mu_{\rm ip}^2 + \frac{32}9\mu_{\rm op}^2 \right)^{-1}.
\label{eq:taued}
\eeq

\section{Plasma excitation and drag}
\label{sec:plasma}

Plasma excitation is the random torquing of dust grains via their interaction with passing ions; plasma drag is the related effect in which a rotating grain spins 
down by transferring its angular momentum to the surrounding plasma.  These processes have been previously computed for uniformly rotating grains in several
papers (\citealt{1993A&A...270..477A}; DL98b; AHD09).

We consider first the excitation in terms of the power spectrum of the electric field at the 
position of the grain.  Then we consider the drag, which is determined using detailed balance arguments.  Finally, we combine this with the analysis of ion 
trajectories by AHD09 to obtain the plasma $F$ and $G$ coefficients.

\subsection{Excitation in terms of electric field power spectrum}

The (nutation angle dependent) plasma excitation coefficient is given by the usual Fokker-Planck rule,
\beq
I_3^2E_\parallel(\Omega, \theta) \Delta t = \langle \Delta L_z^2 \rangle.
\eeq 
This may be evaluated to first order in perturbation theory
by noting that the change in $z$ angular momentum in time $\Delta t$ is equal to the integral of the dipole torque,
\beq
\Delta L_z = \int_0^{\Delta t} (\mu_xE_y-\mu_yE_x) \rmd t,
\eeq
where ${\bmath E}$ is the ambient electric field.  In terms of the rotation matrix ${\mathbfss U}$,
\barr
\Delta L_z &=& \mu_{\rm ip}\int_0^{\Delta t} (U_{xi}E_y-U_{yi}E_x) \rmd t
\nonumber \\ &&
  + \mu_{\rm op}\int_0^{\Delta t} (U_{xk}E_y-U_{yk}E_x) \rmd t.
\earr
Then the excitation coefficient is:
\barr
I_3^2E_\parallel(\Omega, \theta) &=& \frac1{\Delta t} \Big\langle  
\mu_{\rm ip}^2 \int_0^{\Delta t} (U_{xi}E_y-U_{yi}E_x)_t \rmd t
\nonumber \\ && \times
 \int_0^{\Delta t} (U_{xi}E_y-U_{yi}E_x)_{t'} \rmd t'
\nonumber \\ &&
+ \mu_{\rm op}^2 \int_0^{\Delta t} (U_{xk}E_y-U_{yk}E_x)_t \rmd t
\nonumber \\ && \times
 \int_0^{\Delta t} (U_{xk}E_y-U_{yk}E_x)_{t'} \rmd t'
\nonumber \\ &&
+ 2\mu_{\rm ip}\mu_{\rm op} \int_0^{\Delta t} (U_{xi}E_y-U_{yi}E_x)_t \rmd t
\nonumber \\ && \times
 \int_0^{\Delta t} (U_{xk}E_y-U_{yk}E_x)_{t'} \rmd t'
\label{eq:ep1}
\Big\rangle.
\earr
To simplify this, we need to change variables to $\tau=t-t'$ and define the inertial frame electric field correlation function by
\beq 
C_E(\tau) = \langle E_x(t) E_x(t') \rangle = \langle E_y(t) E_y(t') \rangle;
\eeq
the $xx$ and $yy$ correlation functions are equal by isotropy of the plasma, and the mixed components are
uncorrelated, e.g. $\langle E_y(t) E_x(t') \rangle = 0$.  We further assume that $C_E(\tau)\rightarrow 0$ at sufficiently long lag times $\tau$, which is appropriate 
for a thermalized isotropic plasma.  Then if $\Delta t$ is long compared to the decorrelation time (as required for the Fokker-Planck equation to be valid), 
Eq.~(\ref{eq:ep1}) simplifies to
\barr
I_3^2E_\parallel(\Omega, \theta) \!\!&=&\!\! \int_{-\infty}^\infty \rmd\tau\,C_E(\tau) \times \nonumber \\&&
\Bigl\{
\mu_{\rm ip}^2 \langle U_{xi}(t)U_{xi}(t')+U_{yi}(t)U_{yi}(t')\rangle
\nonumber \\ &&
+ \mu_{\rm op}^2 \langle U_{xk}(t)U_{xk}(t')+U_{yk}(t)U_{yk}(t')\rangle
\nonumber \\ &&
+ 2\mu_{\rm ip}\mu_{\rm op} \langle U_{xi}(t)U_{xk}(t')+U_{yi}(t)U_{yk}(t')\rangle
\Bigr\}.
\nonumber \\ &&
\label{eq:ep2}
\earr
Note that to obtain this equation we used the fact that to lowest order, the electric field and the grain orientation are independent, so expressions of the type $\langle E_y(t) E_y(t') U_{xi}(t) U_{xi}(t') \rangle$ can be factored into $\langle E_y(t) E_y(t')\rangle \langle U_{xi}(t) U_{xi}(t') \rangle$.

We now perform the angle ($\phi,\psi$) averages of the correlation functions of the ${\mathbfss U}$ matrix elements using their explicit expressions from 
Eq.~(\ref{eq:U}); for example,
\barr
\langle U_{xk}(t)U_{xk}(t') \rangle &=&
  \sin^2\theta \langle \sin\phi(t) \sin\phi(t') \rangle
\nonumber \\
&=& \sin^2\theta
  \langle\sin\phi(t) \sin[\phi(t)-\dot\phi\tau]\rangle
\nonumber \\
&=& \frac12\sin^2\theta\cos(\dot\phi\tau).
\earr
These simplifications give
\barr
I_3^2E_\parallel(\Omega, \theta) &=& \mu_{\rm ip}^2 \int_{-\infty}^\infty \rmd\tau\,C_E(\tau)
\nonumber \\ && \times
\Bigl\{
\frac{(1-\cos{\theta})^2}{4}\cos[(\dot\phi-\dot\psi)\tau]
\nonumber \\ &&+
\frac{(1+\cos{\theta})^2}{4}\cos[(\dot\phi+\dot\psi)\tau]
\Bigr\}
\nonumber \\ &&
+ \mu_{\rm op}^2 \int\rmd\tau\,C_E(\tau) \sin^2\theta\cos(\dot\phi\tau).
\label{eq:ep3}
\earr

A further simplification can be achieved by switching from the electric field correlation function to its power spectrum, which is easier to compute.  The power 
spectrum 
$P_E(f)$ at frequency $f$ is related to the correlation function via
\beq
\int_{-\infty}^\infty C_E(\tau) \cos{\omega \tau} \ \rmd\tau = P_E\left(\frac{\omega}{2\pi}\right).
\eeq
This reduces Eq.~(\ref{eq:ep3}) to a simple sum,
\barr
I_3^2E_\parallel(\Omega, \theta) &=& \mu_{\rm ip}^2  
\Bigl\{
\frac{(1-\cos{\theta})^2}{4}P_E\left(\frac{\dot\phi-\dot\psi}{2\pi}\right)
\nonumber \\ &&+
\frac{(1+\cos{\theta})^2}{4}P_E\left(\frac{\dot\phi+\dot\psi}{2\pi}\right)
\Bigr\}
\nonumber \\ &&
+ \mu_{\rm op}^2 \sin^2\theta\,P_E\left(\frac{\dot\phi}{2\pi}\right).
\label{eq:ExcJ}
\earr
The excitation coefficient $E_{\parallel}(\Omega)$ used in Eq. (\ref{eq:FP}) can then be obtained by performing the average over nutation angles. 

\subsection{Plasma drag}

The evaluation of the plasma drag is more complicated.  In principle, it is a result of second-order perturbation theory: the dipole moment of the grain modifies the 
trajectories of passing ions, and then the modified charge distribution exerts a torque on the grain with nonzero expectation value.\footnote{For the same reason, 
plasma drag can be thought of as the result of emission of plasma ``waves'' whose amplitude is proportional to $\mu$ and hence whose angular momentum is proportional 
to $\mu^2$ \citep{2002ApJ...568..232R}.}  However, a much simpler method 
of evaluating the plasma drag is to use the principle of detailed balance to relate the rate of small changes in $L$ and $\theta$ to the rate of inverse changes.  
This method works in four stages: first, we need to obtain the diffusion tensor due to plasma drag in $(J,K)$ space (ignoring the thermal spikes); we need to relate 
the damping rate $\langle\Delta J\rangle$ to the diffusion tensor; and then we need 
to express $\tilde D$ in terms of these coefficients.  Finally we perform the average over nutation angles (or equivalently, over $K$ at fixed $J$).

\subsubsection{Diffusion tensor}

The rate of diffusion of a grain in $(J,K)$ space due to plasma excitation is described by a $2\times 2$ symmetric diffusion matrix.  We have already computed the 
component associated with $J$:
\beq
E_{JJ} \equiv \frac{\rmd\langle\Delta J^2\rangle}{\rmd t} = \frac{I_3^2}{\hbar^2}E_\parallel(\Omega, \theta).
\label{eq:ejj}
\eeq
There are also the other components:
\beq
E_{JK} \equiv \frac{\rmd\langle\Delta J\Delta K\rangle}{\rmd t}
\label{eq:ejk}
\eeq
and
\beq
E_{KK} \equiv \frac{\rmd\langle\Delta K^2\rangle}{\rmd t}.
\eeq

We may compute $E_{JK}$ by methods similar to those used to obtain $E_{JJ}$.  The change $\hbar\Delta K$ in the projection of the angular momentum onto the grain 
$\hat{\bmath k}$-axis is equal to the integral of the projection of the torque onto the $\hat{\bmath k}$-axis\footnote{In the second equality here, we have used the 
triple product identity ${\bmath a}\cdot({\bmath b}\times{\bmath c})={\bmath c}\cdot({\bmath a}\times{\bmath b})$.},
\beq
\hbar\Delta K = \int \hat{\bmath k}\cdot(\bmath\mu\times{\bmath E}) \,\rmd t  
 = \int \mu_{\rm ip}{\bmath E}\cdot\hat{\bmath j} \, \rmd t.
\eeq 
The evaluation of Eq.~(\ref{eq:ejk}) gives
\barr
E_{JK} &=& \frac{\mu_{\rm ip}^2}{4\hbar^2}
\Bigl[ (1+\cos{\theta})^2 P_E\left(\frac{\dot\phi + \dot\psi}{2 \pi}\right)
\nonumber \\ &&
 - (1-\cos{\theta})^2 P_E\left(\frac{\dot\phi - \dot\psi}{2\pi}\right) \Bigr].
\label{eq:ejk1}
\earr
We note that $E_{JK}(J,-K)=-E_{JK}(J,K)$ since the two terms in brackets are switched (recall that if $K\rightarrow-K$ then $\theta\rightarrow\pi-\theta$ and $\dot\psi\rightarrow-\dot\psi$).

A similar technique could also be used to compute $E_{KK}$; however
we will not need $E_{KK}$ in our analysis because this does not enter into the equations for $D_J$.

\subsubsection{Relation to drag}

The key to computing plasma drag is the principle of detailed balance.  We note that in true thermal equilibrium with the plasma, and in the absence of thermal spikes
redistributing $K$
(i.e. we consider only plasma interactions as a mechanism of changing $J$ and $K$), the probability of being in the $(J,K)$ rotational level is
\barr
P(J,K)\!\! &\propto& \!\! (2J+1)\exp\left[\frac{-\hbar^2J(J+1)}{2I_1kT}\right]
\nonumber \\ && \times
 \exp\left[\frac{\hbar^2(I_1^{-1}-I_3^{-1})K^2}{2kT}\right],  
\earr
with the factor $2J+1$ representing the $M$-sublevel degeneracy.  

We define $\Gamma_{J,K \rightarrow J', K'}$ to be the rate at which dust grains in the $(J,K)$ quantum state transition to the $(J',K')$ state due to plasma 
excitation.  We further define the quantum number changes $\Delta J=J'-J$ and $\Delta K=K'-K$, and the mean values $J_*=(J+J')/2$ and $K_*=(K+K')/2$.
The principle of detailed balance tells us that
\beq  
\Gamma_{J,K \rightarrow J', K'} P(J,K) = \Gamma_{J',K' \rightarrow J, K} P(J',K').
\eeq
Assuming (as appropriate for the Fokker-Planck approximation) that $|\Delta J|,|\Delta K|\ll J$, we find
\barr
\frac{\Gamma_{J,K \rightarrow J', K'}}{\Gamma_{J',K' \rightarrow J, K}}
\!\!&=&\!\! \frac{2J'+1}{2J+1} \exp\left[\frac{-\hbar^2[ J'(J'+1) - J(J+1)]}{2I_1kT}\right]
\nonumber \\ && \times
\exp\left[\frac{\hbar^2(I_1^{-1}-I_3^{-1})}{2kT}(K'{^2}-K^2)\right] 
\nonumber \\ &\approx&   
1 + \frac{\Delta J}J - \frac{\hbar^2 J\Delta J}{I_1kT}
\nonumber \\ && 
 + \frac{\hbar^2(I_1^{-1}-I_3^{-1}) K\Delta K}{kT}.  
\earr
We then define the symmetrized rate,
\beq
S_{\Delta J,\Delta K}(J_*,K_*) = \frac{\Gamma_{J,K \rightarrow J', K'} + \Gamma_{J',K'\rightarrow J,K}}2,
\eeq
defined at either integer or half-integer values of the arguments depending on whether $\Delta J$ and $\Delta K$ are even or odd.  The rate $S$ is symmetric in the 
sense that $S_{\Delta J,\Delta K}(J_*,K_*)=S_{-\Delta J,-\Delta_K}(J_*,K_*)$.  Then
\barr
\Gamma_{J,K \rightarrow J', K'} &=& 
S_{\Delta J,\Delta K}( J_*,K_*)
\Bigl[ 1 + \frac{\Delta J}{2J} - \frac{\hbar^2 J\Delta J}{2 I_1kT}  
\nonumber \\ &&
 + \frac{\hbar^2(I_1^{-1}-I_3^{-1}) K\Delta K}{2kT} \Bigr];
\earr
Taylor-expanding $S$ and keeping only terms first order in $\Delta J$ and $\Delta K$ gives
\barr
\Gamma_{J,K \rightarrow J', K'} &=& S_{\Delta J,\Delta K}(J,K)\Bigl[ 1 + \frac{\Delta J}{2J} - \frac{\hbar^2 J\Delta J}{2I_1kT}
\nonumber \\ &&
 + \frac{\hbar^2(I_1^{-1}-I_3^{-1}) K\Delta K}{2kT} \Bigr]
\nonumber \\ && 
+ \frac{\Delta J}2\partial_JS_{\Delta J,\Delta K}(J,K)
\nonumber \\ && 
+ \frac{\Delta K}2\partial_KS_{\Delta J,\Delta K}(J,K).
\label{eq:GS}
\earr

We may now relate the excitation rates to the symmetrized rate function.  Inspection of Eq.~(\ref{eq:ejj}) gives
\barr
E_{JJ}(J,K) &=& \sum_{\Delta J\Delta K} \Delta J^2 \Gamma_{J,K \rightarrow J', K'}
\nonumber \\ &=& \sum_{\Delta J\Delta K} \Delta J^2 S_{\Delta J,\Delta K}(J,K),
\earr
and similarly for $E_{JK}$ and $E_{KK}$.  We may then investigate the mean rate of change of $J$:
\beq
\frac{\rmd\langle\Delta J\rangle}{\rmd t} = \sum_{\Delta J\Delta K} \Delta J\; \Gamma_{J,K \rightarrow J', K'}.
\eeq
Here the contributions from $\Delta J,\Delta K$ and $-\Delta J,-\Delta K$ nearly cancel.  They differ only due to the presence of first-order terms (in $\Delta 
J,\Delta K$) in Eq.~(\ref{eq:GS}); these give
\barr
\frac{\rmd\langle\Delta J\rangle}{\rmd t} \!\! &=& \sum_{\Delta J\Delta K} \Delta J \Bigl[
\Bigl( \frac{\Delta J}{2J} - \frac{\hbar^2 J\Delta J}{2I_1kT}
\nonumber \\ &&
+ \frac{\hbar^2 (I_1^{-1}-I_3^{-1})K\Delta K}{2kT}
\Bigr) S_{\Delta J,\Delta K}
\nonumber \\ &&
+ \frac{\Delta J}2\partial_JS_{\Delta J,\Delta K}(J,K)
\nonumber \\ &&
+ \frac{\Delta K}2\partial_KS_{\Delta J,\Delta K}(J,K)
\Bigr]
\nonumber \\
&=&
\left(\frac1{2J} - \frac{\hbar^2J}{2I_1kT}\right)E_{JJ}
+ \frac{\hbar^2 (I_1^{-1}-I_3^{-1})K}{2kT}E_{JK}   
\nonumber \\ &&
+ \frac12\frac{\partial E_{JJ}}{\partial J}
+ \frac12\frac{\partial E_{JK}}{\partial K}.
\label{eq:dj}
\earr
We thus arrive at the remarkable result that the rate of loss of angular momentum due to plasma drag is expressible in terms of $E_{JJ}$ and $E_{JK}$.  
Equation~(\ref{eq:dj}) is the closest that we come to a standard fluctuation-dissipation relation.

\subsubsection{Computation of the drag coefficient $\tilde D$}

In order to continue, we recall that we ultimately need the function $\tilde D(\Omega)$, which first requires us to find $D(\Omega, \theta)$ and its average over nutation angles $D(\Omega)$. 
We recall that
\beq
D(\Omega, \theta) = - \frac{\rmd\langle \Delta \bOmega \rangle}{\rmd t}\cdot \hat{\bmath e}_{\bOmega}.
\eeq
This is {\em not} the same as 
\beq
- \frac{\rmd\langle \Delta \Omega \rangle}{\rmd t} = - \frac{\hbar}{I_3}\frac{\rmd\langle \Delta J \rangle}{\rmd t},
\eeq
where ``$\Delta \Omega$'' is understood as $\Delta |\bOmega|$ and the last equality holds in the large $J$ limit. These two quantities are however related:
\barr
\Delta \Omega &=&  |\bOmega + \Delta \bOmega| - |\bOmega| \nonumber \\
&=& \sqrt{\Omega^2 + 2 \Delta \bOmega \cdot \bOmega + (\Delta \bOmega)^2} - \Omega \label{Delta Omega} \nonumber \\
&=& \Omega \Bigl[\frac{\Delta \bOmega \cdot \bOmega }{\Omega^2} + \frac{(\Delta \bOmega)^2}{2 \Omega^2} - \frac{1}{8}\left(\frac{2 \Delta \bOmega \cdot 
\bOmega}{\Omega^2}\right)^2
\nonumber \\ &&
 + \mathcal O\left(\frac{\Delta \bOmega}{\Omega}\right)^3\Bigr].
\earr
Averaging and taking the time derivative implies
\beq
\frac{\rmd \langle \Delta \Omega \rangle}{\rmd t} = \frac{\rmd\langle \Delta \bOmega \rangle}{\rmd t}\cdot \hat{\bmath e}_{\bOmega} + \frac{E_{\bot}(\Omega, \theta)}{\Omega} ,
\eeq
where the parallel part of the excitation was cancelled by the third term in Eq.(\ref{Delta Omega}).
Solving for $D$ then gives:
\beq
D(\Omega, \theta) = -\frac\hbar{I_3}\frac{\rmd\langle\Delta J\rangle}{\rmd t} + \frac{E_\perp(\Omega, \theta)}{\Omega}, 
\eeq
and averaging over the nutation angle, or equivalently over $K$, gives the coefficient used in the Fokker-Planck equation:
\beq
D(\Omega) = -\frac\hbar{I_3}\Big{\langle}\frac{\rmd\langle\Delta J\rangle}{\rmd t}\Big{\rangle}_{K} + \frac{E_\perp(\Omega)}{\Omega}. 
\eeq
The modified damping coefficient $\tilde D(\Omega)$ of Eq.~(\ref{eq:Dtilde}), is then
\beq
\tilde D = -\frac\hbar{I_3}\Big{\langle}\frac{\rmd\langle\Delta J\rangle}{\rmd t} \Big{\rangle}_{K}+ \frac{E_\parallel}{\Omega} + \frac12\frac{\rmd E_\parallel}{\rmd\Omega}, 
\eeq
or in terms of $J$, and using the averaged Eq.~(\ref{eq:ejj}) $E_{\parallel} = \frac{\hbar^2}{I_3^2}\langle E_{JJ}\rangle_{K}$, we have 
\beq
\frac{I_3}\hbar\tilde D = -\Big{\langle}\frac{\rmd\langle\Delta J\rangle}{\rmd t}\Big{\rangle}_{K} + \frac{\langle E_{JJ}\rangle_K }{J} + \frac12\frac{\rmd}{\rmd J} \langle E_{JJ} \rangle_K. 
\label{eq:plasma-D-master.1}
\eeq
Here the averages are taken over the nutation quantum number $K$.
It is critical to note here that $\rmd/\rmd J$ is a {\em total} derivative, i.e., the averaging over $K$ is 
understood to take place {\em before} the differentiation.  This is because in the definition (Eq.~\ref{eq:Dtilde}), $\tilde D$ is ultimately constructed out of drift 
and diffusion coefficients ${\bmath D}$ and ${\mathbfss E}$ that have already been $K$-averaged.
Thus we {\em cannot} replace the last term with the average of a partial derivative, $\langle\partial E_{JJ}/\partial J\rangle_K$.

Equation~(\ref{eq:plasma-D-master.1}) may be simplified by plugging in Eq.~(\ref{eq:dj}):
\barr
\frac{I_3}\hbar\tilde D &=&  \left(\frac1{2J}+\frac{\hbar^2J}{2I_1kT}\right)\langle E_{JJ}\rangle_K
\nonumber \\ &&
- \frac{\hbar^2 (I_1^{-1}-I_3^{-1})}{2kT} \langle K E_{JK}\rangle_K
- \frac12\Bigl\langle\frac{\partial E_{JJ}}{\partial J}\Bigr\rangle_K
\nonumber \\ &&
- \frac12\Bigl\langle\frac{\partial E_{JK}}{\partial K}\Bigr\rangle_K
+ \frac12\frac{\rmd}{\rmd J} \langle E_{JJ} \rangle_K.
\label{eq:plasma-D-master}
\earr

\subsubsection{Nutation angle average}

Our final step in the above analysis is to perform the average over nutation states $K$.  We would like to express Eq.~(\ref{eq:plasma-D-master}) in a form that does 
not contain any derivatives, since the latter tend to be numerically unstable.  We begin by making the replacement:
\beq
\langle\rangle_K \rightarrow \frac1{2J}\int_{-J}^J \rmd K,
\eeq
valid for large values of $J$ (i.e. the classical regime).  Each of the three derivative-containing terms in Eq.~(\ref{eq:plasma-D-master}) then simplifies.  For 
example,
\barr
\left\langle\frac{\partial E_{JK}}{\partial K}\right\rangle_K &=&
\frac1{2J}\int_{-J}^J\frac{\partial E_{JK}}{\partial K} \rmd K
\nonumber \\
&=& \frac{E_{JK}(J,J)-E_{JK}(J,-J)}{2J}.
\earr
The last term simplifies as well:
\barr
\frac{\rmd}{\rmd J} \langle E_{JJ} \rangle_K 
\!\!&=&
\frac{\rmd}{\rmd J} \left(\frac1{2J}\int_{-J}^J E_{JJ} \rmd K \right)
\nonumber \\ &=&
-\frac1{2J^2}\int_{-J}^J E_{JJ} \rmd K + \frac1{2J}\int_{-J}^J\frac{\partial E_{JJ}}{\partial J} \rmd K
\nonumber \\ &&
+ \frac{E_{JJ}(J,J) + E_{JJ}(J,-J)}{2J}.
\earr
By symmetry under change of sign of $K$ (i.e. $\theta\leftrightarrow\pi-\theta$), we have $E_{JK}(J,-J)=-E_{JK}(J,J)$ and $E_{JJ}(J,-J)=E_{JJ}(J,J)$.  
Also, inspection of Eqs.~(\ref{eq:ExcJ}) and (\ref{eq:ejk1}) in the $K=J$ ($\theta=0$) case shows that\footnote{Although we use the plasma drag calculation to prove it, this is a general result.
If we start at $K=J$ then a small change $\Delta J-\Delta K=J-K\approx J\Delta\theta^2/2$.  Therefore the combination of diffusion coefficients $E_{JJ}-E_{JK}$ evaluates to $\langle\Delta J(\Delta 
J-\Delta K)\rangle/\Delta t = J\langle\Delta J\Delta\theta^2\rangle/\Delta t$.  But diffusion is a $\Delta t^{1/2}$ process so $\langle\Delta J\Delta\theta^2\rangle$ is at least of order $\Delta 
t^{3/2}$.  Hence, taking the limit as $\Delta t\rightarrow 0^+$, $E_{JJ}-E_{JK}$ vanishes.}
$E_{JJ}(J,J)=E_{JK}(J,J)$.
Substituting these results into Eq.~(\ref{eq:plasma-D-master}), we find a mass cancellation resulting in
\barr
\frac{I_3}\hbar\tilde D &=&
\frac{\hbar^2}{4I_1kT} \int_{-J}^J E_{JJ} \rmd K
\nonumber \\ &&
- \frac{\hbar^2 (I_1^{-1}-I_3^{-1})}{4JkT}\int_{-J}^JKE_{JK}\rmd K.
\label{eq:plasma-D}
\earr

The detailed balance-derived drag coefficient can be generalized for any bath at temperature $T_X$, and can be written in the form:
\beq
\tilde D = \frac{I_3 \Omega}{2 k T_X} \left[ \frac{I_3}{I_1}E_{\parallel}  -  \left(\frac{I_3}{I_1} - 1\right) \Bigl\langle \frac{K}{J} \frac{\hbar^2 E_{JK}}{I_3^2}\Bigr\rangle_K\right].
\label{eq:detailed.balance}
\eeq
This should be compared to the fluctuation-dissipation theorem, $\tilde D = I_3 \Omega E_{\parallel}/2 k T_X$, valid for a grain rotating around its axis of greatest 
inertia.  In particular, Eq.~(\ref{eq:detailed.balance}) reduces to the fluctuation-dissipation theorem in the limit of spherical grain ($I_3/I_1=1$).

\subsection{Computation of $G$ and $F$ coefficients}

We are finally ready to construct formulas for the $G$ and $F$ coefficients.  It is most convenient to express these in terms of the
AHD09 excitation coefficients $G_{\rm p,\rm AHD}(\Omega)$, which of course have already been calculated.  Recall that on account of the 
fluctuation-dissipation theorem we had
$F_{\rm p,\rm AHD}(\Omega)=G_{\rm p,\rm AHD}(\Omega)$.  In all cases, we set $I_1=\frac12I_3$. 

\subsubsection{$G$ coefficient}

We recall that the AHD09 excitation coefficient was derived by assuming $\theta=0$, in which case Eq.~(\ref{eq:ExcJ}) reduces to
\beq
I_3^2E_\parallel(\Omega;\theta=0) = \mu_{\rm ip}^2 P_E\left(\frac\Omega{2\pi}\right).
\eeq
The plasma excitation rate, using Eq.~(\ref{eq:G}), is then
\beq
G_{\rm p,AHD}(\Omega) = \frac{\tau_{\rm H}}{2I_3kT}\mu_{\rm ip}^2 P_E\left(\frac\Omega{2\pi}\right).
\eeq
This allows us to express the electric field power spectrum $P_E(f)$ in terms of the AHD09 excitation coefficients:
\beq
P_E(f) = \frac{2I_3kT}{\tau_{\rm H}\mu_{\rm ip}^2} G_{\rm p,AHD}(2\pi f).
\label{eq:pef}
\eeq

We may now use Eq.~(\ref{eq:ExcJ}) to obtain the plasma excitation rate for general $\theta$.  Recalling that $\dot\phi=2\Omega$ and $\dot\psi=-\Omega\cos\theta$, we 
find
\barr
G_{\rm p}(\Omega, \theta) &=& \frac{(1-\cos\theta)^2}{4} G_{\rm p,AHD}[(2+\cos\theta)\Omega] 
\nonumber \\ &&
 + \frac{(1+\cos\theta)^2}{4} G_{\rm p,AHD}[(2-\cos\theta)\Omega]
\nonumber \\
&& + \frac{\mu_{\rm op}^2}{\mu_{\rm ip}^2}\sin^2\theta\,G_{\rm p,AHD}(2\Omega).
\earr
We now average over values of $\cos\theta$ between $-1$ and $+1$.  The first two terms give identical contributions, and the last one simplifies using 
$\langle\sin^2\theta\rangle=\frac23$.  Thus
\barr
G_{\rm p}(\Omega) &=& \frac14 \int_\Omega^{3\Omega} \left( 3 -\frac{\omega}\Omega \right)^2 G_{\rm p,AHD}(\omega)\, \frac{\rmd\omega}\Omega
\nonumber \\
&& + \frac{2\mu_{\rm op}^2}{3\mu_{\rm ip}^2}\,G_{\rm p,AHD}(2\Omega). 
\label{eq:Gp}
\earr

\subsubsection{$F$ coefficient}

A similar technique works for the drag coefficient.  We substitute Eq.~(\ref{eq:plasma-D}) for $\tilde D$ into Eq.~(\ref{eq:F}) to obtain an expression for $F_{\rm 
p}(\Omega)$.  This in turn depends on the excitation coefficients $E_{JJ}$ [from Eqs.~(\ref{eq:ejj}) and (\ref{eq:ExcJ})] and $E_{JK}$ [from Eq.~(\ref{eq:ejk1})].  
This leads to an expression involving electric field power spectra, which we transform into AHD09 coefficients using Eq.~(\ref{eq:pef}).
Converting the integrals over $K$ to integrals over $\cos\theta=K/J$ and using $\hbar J=I_3\Omega$, we find a mass cancellation of prefactors, giving
\barr
F_{\rm p}(\Omega) \!\!\!\! &=& \!\!\!\! 2 G_{\rm p}(\Omega) 
\nonumber \\ && \!\!\!\!
- \frac12 \int_{-1}^1 \Bigl\{ 
\frac{(1+\cos\theta)^2}4 G_{\rm p,AHD}[(2-\cos\theta)\Omega]\nonumber \\ 
&&\!\!\!\!
- \frac{(1-\cos\theta)^2}{4} G_{\rm p,AHD}[(2+\cos\theta)\Omega]\Bigr\}
\nonumber \\ &&\!\!\!\!
 \cos \theta \rmd \cos \theta.
\earr
This simplifies to
\barr
F_{\rm p}(\Omega) \!\! &=& \!\! \frac14 \int_\Omega^{3\Omega}\frac{\omega}{\Omega} 
\left( 3 -\frac{\omega}\Omega \right)^2
G_{\rm p,AHD}(\omega) \frac{\rmd\omega}\Omega
\nonumber \\
&&\!\! + \frac{4\mu_{\rm op}^2}{3\mu_{\rm ip}^2} G_{\rm p,AHD}(2\Omega).
\label{eq:Fp}
\earr

Note that due to the $K$-averaging there is no longer a definite relation between $F$ and $G$.  However, since $G_{\rm p,AHD}(\omega)>0$ for all $\omega$, we find 
the inequality
\beq
1 < \frac{F_{\rm p}(\Omega)}{G_{\rm p}(\Omega)} < 3.
\eeq


The calculation of $G_{\rm p, \rm AHD}(\omega)$ is one of the most time-consuming parts of {\sc SpDust}, but it varies slowly as a function of frequency and is only 
required over a range of a factor of 3 in frequency ($\Omega<\omega<3\Omega$).  Thus we 
implement it with an approximate integrator as described in Appendix~\ref{app:Gp}.

\begin{figure} 
\includegraphics[width = 3.2in]{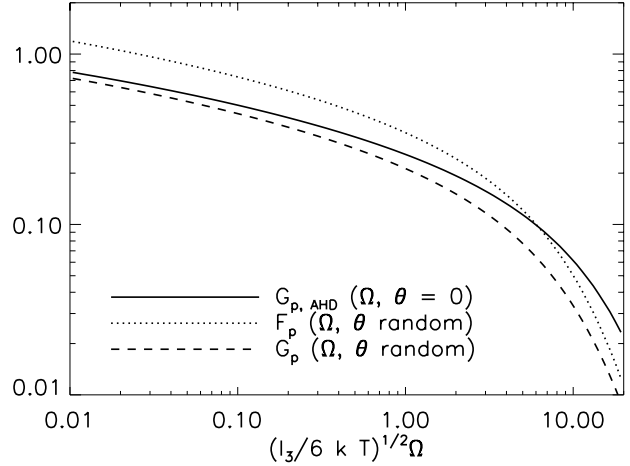}
\caption{Plasma excitation and drag dimensionless coefficients, for $N_{\rm C}=54$ in the WIM.}
\label{fig:FpGp}
\end{figure}

We show the plasma excitation and drag coefficient for a grain containing $N_{\rm C}=54$ carbon atoms (equivalent radius $a \approx 5$\AA), in WIM conditions (as 
defined in DL98b) in Fig.~\ref{fig:FpGp}.

\section{Infrared excitation and damping}
\label{sec:ir}

Another major spin-up/down mechanism for the smallest grains is the emission of infrared photons during thermal spikes.  Here we consider the excitation and damping 
due to these spikes.

The excitation rate is doubled from the AHD09 treatment for {\em all} grains (spherical or not) due to a previous error associated with the 
emitted photon angular momentum (the angular momentum carried away by a photon is $\sqrt2\,\hbar$ rather than just $\hbar$).  This is the only modification in this 
paper that applies to spherical as well as disc-like grains.

\subsection{Excitation}
\label{ss:irex}

Infrared excitation is the random change in angular momentum resulting from the fact that each infrared photon emitted by the grain carries away some angular 
momentum.  In the previous analyses of DL98b and AHD09, it was assumed that the resulting change in angular momentum had 
variance $\langle \Delta{\bmath L}^2\rangle=\hbar^2$ (since a photon carries one quantum of angular momentum), or on one axis $\langle \Delta 
L_z^2\rangle=\frac13\hbar^2$.  In fact, the excitation is twice this as can be seen from either of the following arguments:
\begin{itemize}
\item
An electric dipole photon has angular momentum quantum number $j=1$, so if it carries off angular momentum $-\Delta{\bmath L}$ (thereby imparting $\Delta{\bmath L}$ 
to the grain via back-reaction), we have $\Delta{\bmath L}^2 = j(j+1)\hbar^2 = 2\hbar^2$.
\item
The $z$-component of the angular momentum of the photon is $\Delta L_z=-m\hbar$, where $m\in\{-1,0,1\}$ is the azimuthal quantum number of the emitted photon.  Since 
these three possibilities are equally likely for an isotropically oriented grain, we see that $\langle\Delta L_z^2\rangle = \frac23\hbar^2$.
\end{itemize}
Both of these arguments show that the infrared excitation $G_{\rm IR}$ is twice that reported in AHD09, i.e.
\beq
G_{\rm IR} = \frac {h\tau_{\rm H}}{3\pi I_3kT} \int_0^\infty \frac{F_\nu }\nu \,\rmd\nu,
\label{eq:GIR}
\eeq
where $F_\nu$ is the spectrum of infrared radiation emitted by the grain (in e.g. erg$\,$s$^{-1}\,$Hz$^{-1}\,$sr$^{-1}$).

The correct excitation rate {\em was} included in \citet{2010A&A...509A..12Y}, however their formalism is quite different (e.g. they use $J$ rather than $\Omega$ as 
the independent variable) and so the discrepancy appears to have not been noticed previously.

\subsection{Damping}
\label{ss:irdamp}

We next consider the infrared damping rate, which arises due to slight preferential emission of positive over negative angular momentum photons from a rotating grain.  
A classical model of the effect can be constructed by considering oscillators either in the plane of the grain or out of the plane.  The torque from an isotropic 
distribution of oscillators would correspond to adding $\frac23$ of the in-plane and $\frac13$ of the out-of-plane result.

We consider an oscillating dipole ${\bmath p}$ with angular frequency $\omega = 2\pi\nu$ and amplitude $P$.  In the out-of-plane case, this corresponds to a dipole 
moment:
\barr
{\bmath p} &=& P\sin(\omega t)\,\hat{\bmath k} \nonumber \\
&=& P\sin(\omega t)(\sin\theta\sin\phi,-\sin\theta\cos\phi,\cos\theta).
\earr
The torque on the grain is then
\beq
{\bmath T} = -\frac2{3c^3}\langle\dot{\bmath p}\times\ddot{\bmath p}\rangle,
\eeq
where the derivatives are taken in the inertial frame.  The evaluation of the time average of the $z$-component of the torque is then a straightforward exercise;
to lowest order in $\dot\phi$, we find
\beq
T_z^{\rm op} = -\frac{\omega^2P^2\sin^2\theta}{c^3}\dot\phi.
\eeq
A similar exercise for an in-plane oscillator gives
\barr
T_z^{\rm ip} &=& -\frac{\omega^2P^2}{4c^3} \Bigl[
(1+\cos\theta)^2(\dot\phi + \dot\psi)
\nonumber \\ &&
+(1-\cos\theta)^2(\dot\phi - \dot\psi)
\Bigr].
\earr
If we average these over nutation angles, we get
\beq
\langle T_z^{\rm op}\rangle_\theta = -\frac{4\omega^2P^2}{3c^3}\Omega
\eeq
and
\beq
\langle T_z^{\rm ip}\rangle_\theta = -\frac{\omega^2P^2}{c^3}\Omega.
\eeq

In order to calculate damping coefficients, we must sum over all the oscillators $P$ that contribute to the infrared emission.
The total power emitted by this dipole is
\beq
4\pi F_\nu = \frac{\omega^4P^2}{3c^3}\delta\left(\nu-\frac\omega{2\pi}\right),
\eeq
so we make the replacement:
\beq
P^2 \rightarrow \int_0^\infty \rmd\nu\,\frac{12\pi c^3}{\omega^4}F_\nu
= \int_0^\infty \rmd\nu\,\frac{3c^3}{\pi\omega^2\nu^2}F_\nu,
\eeq
with $\omega=2\pi\nu$.  The total torque is then
\beq
T = -\int_0^\infty \rmd\nu\,\frac{\Omega}{\pi\nu^2}(3F_\nu^{\rm ip} + 4F_\nu^{\rm op}),
\eeq
where $F_\nu^{\rm ip}$ and $F_\nu^{\rm op}$ are the emission spectra contributed by the in-plane and out-of-plane modes.
The damping coefficient is
\beq
F_{\rm IR} = \frac{\tau_{\rm H}}{\pi I_3} \int_0^\infty \rmd\nu\,\frac{3F_\nu^{\rm ip} + 4F_\nu^{\rm op}}{\nu^2}.
\eeq
We can see that there is very little difference between the in-plane and out-of-plane mode contributions (a factor of $\frac43$).
Assuming the isotropic case where $\frac23$ of the emission 
is in-plane and $\frac13$ is out-of-plane\footnote{In the case of the PAH bands, it is known that some bands correspond to
in-plane vibrations and some to out-of-plane; however given the small difference between the two cases, we have not
tracked them separately.}, we find
\beq
F_{\rm IR} = \frac{10\tau_{\rm H}}{3\pi I_3} \int_0^\infty \frac{F_\nu}{\nu^2}\,\rmd\nu.
\label{eq:FIR}
\eeq
This is $\frac53$ times the AHD09 damping coefficient for spherical grains.

\section{Collisions}
\label{sec:coll}

Collisions of dust grains break down into several cases: the grain may be charged or neutral, and the impactor may be ionized or neutral.  Furthermore,
one must consider not just the angular momentum imparted by the incoming particle, but also how much angular momentum it carries away when it
evaporates.  We denote the damping and excitation rates with subscripts $_{\rm i}$ or $_{\rm n}$ (for ion or neutral impactor) and superscripts
$^{\rm(in)}$ or $^{\rm(ev)}$ for incoming or evaporative contributions.  In the $\theta=0$ case, the grain's geometry is time-stationary in the inertial frame, and incoming particles are equally likely 
to impact
the grain whether they approach on prograde or retrograde trajectories, and hence $F^{\rm(in)}=0$.  For the more general case, there will be a new $F_{\rm n}^{\rm(in)}$ contribution
associated with the fact that the grain can physically crash into passing particles, and this leads to a preference to accrete incoming particles on retrograde orbits.

The general problem is not tractable analytically, so we focus first on the case of neutral impactors on neutral grains.  We then
heuristically extend the calculation to the more general case.


\subsection{Damping rate: neutral grains, neutral impactors}
\label{ss:nnd}

There are two contributions to the damping rate.  The first is the evaporative damping, $F_{\rm coll}^{\rm(ev)}$, which arises because particles evaporating off the grain surface preferentially have 
positive $L_z$.  The second is a new contribution, $F_{\rm coll}^{\rm(in)}$, which arises because a grain rotating around an axis other than a symmetry axis preferentially collides with incoming 
particles of negative $L_z$.  We consider both in turn.  In both cases, we assume the grain to be a convex rigid body whose surface area element is $\rmd S$, whose normal vector is $\hat{\bmath n}$, and whose 
instantaneous angular velocity is $\bomega$. \\[10pt]

\subsubsection{Evaporation}

We suppose that a particle evaporates from position ${\bmath r}$ on the grain surface.  This point has a local surface velocity ${\bmath v}_0=\bomega\times{\bmath r}$.  The local phase-space density of 
particles evaporating from the grain surface is
\beq
f({\bmath r}, {\bmath v}) = K \exp \left[ -\frac{m(\bmath{v} - \bmath{v}_0)^2}{2 k T_{\rm ev}}\right],
\label{eq:evap-f}
\eeq
for ${\bmath v}$ in the half-space:
\beq
{\cal H} \equiv \{ {\bmath v} \in\mathbb R^3: ({\bmath v}-{\bmath v}_0)\cdot\hat{\bmath n}>0 \}.
\eeq
The normalization constant $K$ can be found from the requirement that the rate of collisions per unit area is equal to the rate of evaporation per unit area.  The flux of evaporating particles (in 
particles$\,$cm$^{-2}\,$s$^{-1}$) is obtained by integrating $(\bmath{v} - \bmath{v}_0)\cdot\hat{\bmath n}f$ over ${\cal H}$, giving:
\beq
\frac \pi2 \left( \frac{2kT_{\rm ev}}{m} \right)^2 K = \frac1S \frac{\rmd N_{\rm coll}}{\rmd t}.
\eeq

The angular momentum imparted to the grain by an individual escaping atom is obtained from Newton's third law, $\Delta{\bmath L} = -m{\bmath r}\times{\bmath v}$.  For an ensemble of escaping atoms, we 
should write
\beq
\frac{\rmd\langle \Delta{\bmath L}\rangle}{\rmd t} = m\oint \rmd S \int_{\cal H} \rmd^3{\bmath v}
(-{\bmath r}\times{\bmath v})
[(\bmath{v} - \bmath{v}_0)\cdot\hat{\bmath n}]
f({\bmath r}, {\bmath v}).
\eeq
The velocity integral is straightforwardly evaluated using the substitution ${\bmath v}={\bmath v}_0+{\bmath u}$.  The result is
\barr
\frac{\rmd\langle \Delta{\bmath L}\rangle}{\rmd t} \!\!\! &=& \!\!\! mK \oint \rmd S \Bigl[ -({\bmath r}\times{\bmath v}_0) 2\pi \left( \frac{kT_{\rm ev}}{m} \right)^2
\nonumber \\ &&\!\!\! - \int_{{\bmath u}\cdot\hat{\bmath n}>0}\! ({\bmath r}\times{\bmath u})({\bmath u}\cdot\hat{\bmath n}) \rme^{-m u^2/2kT_{\rm ev}} \rmd^3{\bmath u} \Bigr].
\label{eq:drift-ev1}
\earr
The second integral has an integrand even in ${\bmath u}$, so its value is exactly $\frac12$ of the integral extended over all ${\bmath u}\in\mathbb{R}^3$.  The resulting integrand is then a quadratic 
function of the components of ${\bmath u}$ times a Gaussian.  Such integrals are easily evaluated; in this case, the result is proportional to ${\bmath r}\times\hat{\bmath n}$.\footnote{This can be 
seen from symmetry, since the integral must be linear in ${\bmath r}$ and $\hat{\bmath n}$, and has the symmetry of a pseudovector.}  But we know that $\oint {\bmath r}\times\hat{\bmath n}\,\rmd S=0$ 
for any closed surface, so the second integral vanishes.\footnote{This is based on the assumption that the evaporation properties (e.g. $T_{\rm ev}$) are uniform across the grain surface.  If this is 
violated, e.g. by catalytic sites for the formation of H$_2$, then there can be a systematic torque.  This has been previously investigated and found to be negligible for the smallest grains (DL98b).}
Therefore, we keep only the first integral.  Using ${\bmath v}_0=\bomega\times{\bmath r}$, we reduce Eq.~(\ref{eq:drift-ev1}) to
\beq
\frac{\rmd\langle \Delta{\bmath L}\rangle}{\rmd t} = -\frac mS \frac{\rmd N_{\rm coll}}{\rmd t} \oint {\bmath r}\times(\bomega\times{\bmath r})\,\rmd S.
\eeq

In the particular case of a disc-like grain of uniform and infinitesimal thickness, the surface average of ${\bmath r}\times(\bomega\times{\bmath r})$ is the same as 
its volume average, which by inspection is the angular momentum ${\bmath L}$ divided by the grain mass $M$.  Thus,
\beq
\frac{\rmd\langle \Delta{\bmath L}\rangle}{\rmd t} = -\frac mM \frac{\rmd N_{\rm coll}}{\rmd t} {\bmath L},
\eeq
or
\beq
D(\Omega) = \frac mM \frac{\rmd N_{\rm coll}}{\rmd t}.
\eeq
This does not depend on $\theta$, so the evaporation contribution to the damping is not modified from the principal axis case.  The relation
\beq
F^{\rm(ev)}_{\rm n} = \frac{n_{\rm n}}{n_{\rm H}}\sqrt{\frac{m_{\rm n}}{m_{\rm H}}}
\eeq
for neutral atoms impacting neutral grains remains valid.

\subsubsection{Incoming particles}

We now require the angular momentum acquired from incoming particles.  This is actually very similar to the previous calculation, except that the phase space density of incoming atoms has zero net velocity,
\beq
f({\bmath r}, {\bmath v}) = n \left( \frac{m}{2\pi kT} \right)^{3/2} \exp \left( -\frac{mv^2}{2 k T}\right),
\eeq
and the relevant region of velocity space is now the complement of ${\cal H}$, i.e. ${\cal H}^{\rm c}$.
The angular momentum transfer rate is
\beq
\frac{\rmd \langle\Delta {\bmath L}\rangle}{\rmd t} = m \oint \rmd S \int_{{\cal H}^{\rm c}} \rmd^3{\bmath v}\,({\bmath r}\times{\bmath v}) [-({\bmath v}-{\bmath v}_0)\cdot\hat{\bmath n}]f({\bmath r},{\bmath v}).
\eeq
We now Taylor-expand to first order in ${\bmath v}_0$.  The zeroeth-order term (i.e. for ${\bmath v}_0={\bmath 0}$) is proportional to $\oint {\bmath r}\times\hat{\bmath n}\,\rmd S=0$ and vanishes.  
There are two possible contributions to the first-order term.  One arises from the explicit ${\bmath v}_0$ in the integrand.  The other arises from the dependence of the integration region ${\cal 
H}^{\rm c}$ on ${\bmath v}_0$.  That is, to first order in ${\bmath v}_0$,
\barr
\frac{\rmd \langle\Delta {\bmath L}\rangle}{\rmd t} \!\!\!\! &=& \!\!\!\! m \oint \rmd S \Bigl[
\int_{{\cal H}^{\rm c}} \rmd^3{\bmath v}\,({\bmath r}\times{\bmath v})
({\bmath v}_0\cdot\hat{\bmath n})\,f({\bmath r},{\bmath v})
\nonumber \\
&& \!\!\!\! + \int_{\partial{\cal H}^{\rm c}}\!
\rmd^2{\bmath v}\,({\bmath v}_0\cdot\hat{\bmath n})
({\bmath r}\times{\bmath v})
(-{\bmath v}\cdot\hat{\bmath n})\,f({\bmath r},{\bmath v})
\Bigr],
\label{eq:temp-bdy}
\earr
where the integration region ${\cal H}^{\rm c}$ is evaluated at ${\bmath v}_0={\bmath 0}$; $\partial{\cal H}^{\rm c}$ is the boundary of ${\cal H}^{\rm c}$; and
$\rmd^2{\bmath v}$ is the area element on the boundary.  The boundary at nonzero ${\bmath v}_0$ is displaced a distance ${\bmath v}_0\cdot\hat{\bmath n}$, hence the 
inclusion of this factor in the second term, combined with the area element $\rmd^2{\bmath v}$, is the element of volume that is brought inside ${\cal H}^{\rm c}$ due 
to nonzero ${\bmath v}_0$.\footnote{The $+$ sign for this term arises because for ${\bmath v}_0\cdot\hat{\bmath n}>0$, ${\cal H}^{\rm c}$ expands.}  The second term 
can be seen to vanish because ${\bmath v}\cdot\hat{\bmath n}=0$ on the boundary $\partial{\cal H}^{\rm c}$.  Therefore this second term may be dropped.


Using the Maxwellian distribution, we may perform the velocity integral in the first (surviving) term in Eq.~(\ref{eq:temp-bdy}) to get
\beq
\frac{\rmd \langle\Delta {\bmath L}\rangle}{\rmd t} = -n\sqrt{\frac{mkT}{2\pi}} \oint \rmd S \, ({\bmath r}\times\hat{\bmath n})({\bmath v}_0\cdot\hat{\bmath n}).
\eeq
Substituting ${\bmath v}_0=\bomega\times{\bmath r}$, we conclude that
\beq
\frac{\rmd \langle\Delta {\bmath L}\rangle}{\rmd t} = -n\sqrt{\frac{mkT}{2\pi}} \oint \rmd S\,({\bmath r}\times\hat{\bmath n})[({\bmath r}\times\hat{\bmath 
n})\cdot\bomega].
\eeq
The triple product implies that this zero if $\bomega$, ${\bmath r}$, and $\hat{\bmath n}$ are coplanar, which is the case for grains rotating around an axis of 
symmetry.  In our case, however, it is nonzero.  We note that the average value of the dyadic $({\bmath r}\times\hat{\bmath n})({\bmath r}\times\hat{\bmath n})$ over 
a disc is $\frac14R^2(\hat{\bmath i}\hat{\bmath i}+\hat{\bmath j}\hat{\bmath j})$, where $R$ is the disc radius and $\hat{\bmath i}\hat{\bmath i}+\hat{\bmath 
j}\hat{\bmath j}$ is the projector into the plane of the grain.  Therefore,
\beq
\frac{\rmd \langle\Delta {\bmath L}\rangle}{\rmd t} = -n\sqrt{\frac{mkT}{2\pi}}\, \frac14R^2S\bomega_{\rm ip},
\eeq
where $\bomega_{\rm ip}$ is the in-plane part of the instantaneous angular velocity.  It is equal to $\bomega_{\rm ip}={\bmath L}_{\rm ip}/I_1$.  Further using 
$I_1=\frac12I_3$, we find:
\barr
\frac{\rmd \langle\Delta {\bmath L}\rangle}{\rmd t} &=& -n\sqrt{\frac{mkT}{2\pi}}\, \frac{\pi R^4}{I_3}
{\bmath L}_{\rm ip}
\nonumber \\
&=& -\frac{n}{n_{\rm H}}\sqrt{\frac{m}{\mH}} \tau_{\rm H}^{-1}
{\bmath L}_{\rm ip},
\earr
where we have used the definition of $\tau_{\rm H}$ and $a_{\rm cx}^4\equiv\frac38R^4$ in the last line\footnote{The excitation radius $a_{\rm cx}$ is the same as in AHD09 when taking the limit of infinitesimally thin disks (AHD09 assumed disks with a thickness $d = 3.35$ \AA). }.

In order to complete the derivation, we need the mean value of ${\bmath L}_{\rm ip}$ over nutation angles and time.  We note that this mean value must be in the 
direction of ${\bmath L}$, and that
\beq
{\bmath L}_{\rm ip}\cdot{\bmath L} = L_{\rm ip}^2 = L^2\sin^2\theta,
\eeq
which has mean value $\frac23L^2$.  Therefore the mean value of ${\bmath L}_{\rm ip}$ is $\frac23{\bmath L}$ and we find:
\beq
F_{\rm n}^{\rm(in)} = \frac23 \frac{n}{n_{\rm H}}\sqrt{\frac{m}{\mH}}.
\eeq
The total drag coefficient is the sum,
\beq
F_{\rm n}=F_{\rm n}^{\rm(in)}+F_{\rm n}^{\rm(ev)} = \frac53 \frac{n}{n_{\rm H}}\sqrt{\frac{m}{\mH}}.
\label{eq:fnn}
\eeq

\subsection{Excitation rate: neutral grains, neutral impactors}
\label{ss:nne}

We now consider the stochastic change in angular momentum due to collisions with incoming particles.  This excitation rate (unlike the damping rate) can be computed 
at zero grain rotation.  The impact of a particle with velocity ${\bmath v}$ at position ${\bmath r}$ imparts an angular momentum $\Delta{\bmath L}=m{\bmath 
r}\times{\bmath v}$.  The stochastic change in angular momentum along the z-axis can be written as
\beq
\frac{\rmd\langle \Delta L_z^2\rangle}{\rmd t} = m^2 \oint \rmd S \int_{{\cal H}^{\rm c}}\! \rmd^3{\bmath v} \,
[\hat{\bmath z}\cdot({\bmath r}\times{\bmath v})]^2
(-{\bmath v}\cdot\hat{\bmath n}) f({\bmath r},{\bmath v}).
\eeq
The triple product can be cyclically permuted to get
\beq
\frac{\rmd\langle \Delta L_z^2\rangle}{\rmd t} = -m^2 \oint \rmd S \int_{{\cal H}^{\rm c}}\! \rmd^3{\bmath v} \,
({\bmath v}\cdot{\bmath q})^2
({\bmath v}\cdot\hat{\bmath n}) f({\bmath r},{\bmath v}),
\eeq
where ${\bmath q}\equiv\hat{\bmath z}\times{\bmath r}$.  The integration over velocity is a Gaussian times a cubic polynomial over a half-space, which evaluates to
\beq
-\frac1{\sqrt{2\pi}}\,n\left(\frac{kT}m\right)^{3/2} [\,q^2 + ({\bmath q}\cdot\hat{\bmath n})^2\,],
\eeq
so
\beq
\frac{\rmd\langle \Delta L_z^2\rangle}{\rmd t} = \sqrt{\frac{ m (kT)^3}{2\pi}} \,n \oint\rmd S \,[\,q^2 + ({\bmath q}\cdot\hat{\bmath n})^2\,].
\label{eq:dlz2-coll}
\eeq
Now the integrand is a scalar and hence may be evaluated in either inertial or grain-fixed coordinates.  We choose the grain-fixed coordinates.  The nutation angle 
average is then equivalent to averaging over the direction of $\hat{\bmath z}$, which leads to the dyadic relation
\beq
\langle{ \bmath{qq}} \rangle = \frac13 (r^2{\mathbfss 1}-{\bmath{rr}}),
\eeq
where ${\mathbfss 1}$ is the unit dyadic.  This implies that
\beq
\langle q^2 + ({\bmath q}\cdot\hat{\bmath n})^2 \rangle = r^2-\frac13({\bmath r}\cdot\hat{\bmath n})^2.
\eeq
Plugging into Eq.~(\ref{eq:dlz2-coll}) and converting to the $G$-factor gives
\beq
G_{\rm n}^{\rm(in)} =
\frac{n}{n_{\rm H}}\sqrt{\frac{m}{\mH}}\,\frac{3}{16\pi a_{\rm cx}^4}
\oint\rmd S [r^2-\frac13({\bmath r}\cdot\hat{\bmath n})^2].
\eeq
For a disc, the integral is $\pi R^4$ and $a_{\rm cx}^4=\frac38R^4$, so it follows that:
\beq
G_{\rm n}^{\rm(in)} = \frac12
\frac{n}{n_{\rm H}}\sqrt{\frac{m}{\mH}}.
\label{eq:Gnn-in}
\eeq
Thus the collisional excitation rate for incoming particles is the same as it is for the case of the grain rotating around a principal axis of inertia.

The calculation for evaporating particles is the same except that we replace $T\rightarrow T_{\rm ev}$:
\beq
G_{\rm n}^{\rm(ev)} =\frac12
\frac{n}{n_{\rm H}}\sqrt{\frac{m}{\mH}} \frac{T_{\rm ev}}T.
\label{eq:Gnn-ev}
\eeq
Once again, there is no difference from the case of rotation around $I_3$.

\subsection{Excitation and damping: charged grain, neutral impactor}

The case of a charged grain is different from a neutral grain because of the induced dipole attraction between the grain and the atom.  The
interaction potential is given by
\beq
V(r) = -\frac{1}{2} \alpha \frac{Z_g^2 q_e^2}{r^4}.
\eeq
We can solve for the critical separation $r_{\rm c}$ at which the 
induced 
dipole attraction overwhelms the thermal energy of the gas, i.e. where $V(r_{\rm c})=\frac32kT$:
\beq
r_{\rm c} = \sqrt[4]{\frac{Z_g^2q_e^2\alpha}{3kT}} \approx 1.5 \left(\frac{\alpha Z_g^2}{0.67 {\rm\AA}^3}\right)^{1/4} \left(\frac{T}{8000 \ \textrm{K}}\right)^{-1/4}
{\rm\,\AA}.
\eeq

In the cases where $r_{\rm c}\ll a_{\rm cx}$, the induced dipole attraction is a small perturbation and the coefficients $F_{\rm n}^{\rm(in)}$, $G_{\rm n}^{\rm(in)}$, 
$F_{\rm n}^{\rm(ev)}$, and $G_{\rm n}^{\rm(ev)}$ are unchanged from the case of a neutral grain.  On the other hand, if $r_{\rm c}\gg a_{\rm cx}$, then an incoming 
particle is 
certain to impact the grain surface if it passes over the barrier in the effective potential $V_{\rm err}(r)=L_n^2/mr^2+V(r)$, irrespective of the details of the 
asymmetry of the grain (a disc-like grain has a quadrupole moment, but at $r\gg a_{\rm cx}$ the potential is dominated by the monopole charge).  Therefore in this 
alternative case, the shape of the grain is irrelevant, and we should use the AHD09 values for the coefficients $F_{\rm n}^{\rm(in)}$ and $G_{\rm n}^{\rm(in)}$.

While $G_{\rm n}^{\rm(in)}$ is the same in both of our limiting cases ($r_{\rm c}/a_{\rm cx}\gg1$ or $\ll1$), $F_{\rm n}^{\rm(in)}$ is not the same and it is 
necessary to 
interpolate between the two solutions.  We must have $F_{\rm n}^{\rm(in)}/F_{\rm n}\rightarrow 0$ for $r_{\rm c}/a_{\rm cx}\gg 1$ and $F_{\rm n}^{\rm(in)}\rightarrow 
\frac23F_{\rm n,AHD09}$ for $r_{\rm c}/a_{\rm cx}\ll 1$. 
A simple heuristic interpolating function\footnote{Since $F_{\rm n, AHD09} \propto \left(\frac{r_{\rm c}}{a_{\rm cx}}\right)^4$ for $r_{\rm c} \gg a_{\rm cx}$, our heuristic prescription for $F_{\rm n}^{\rm(in)}$ is such that $F_{\rm n}^{\rm(in)} \nrightarrow 0$ in that limit. It is not clear whether $F_{\rm n}^{\rm(in)}$ should tend to zero for $r_{\rm c} \gg a_{\rm cx}$, since although the relative difference in collisional rates between impactors on prograde orbits and retrograde orbits should vanish, the overall collision rate increases because of  electrostatic focusing. However, only the relative contribution $F_{\rm n}^{\rm(in)}/ F_{\rm n}$ matters so this should not be a concern.} is
\beq
F_{\rm n}^{\rm(in)} = \frac23F_{\rm n,AHD09} \left[ 1 + \left( \frac{r_{\rm c}}{a_{\rm cx}} \right)^2 \right]^{-1}.
\eeq
In diffuses phases of the ISM, collisions with neutral impactors are in general not the dominant rotational damping mechanism (see for example Fig.~4 of DL98b). The 
exact shape of the interpolation function is therefore irrelevant in these cases. In very specific environments though, for example in extremely dense PDRs, and for low enough values of the dipole moment, collisions of neutral impactors on charged grains may dominate the rotational damping. If this is the case, one should be aware that $F_{\rm n}$ is uncertain in the region $r_{\rm c} \sim a_{\rm cx}$ and that this uncertainty will propagate on the resulting spectrum, as $\nu_{\rm peak} \propto \sqrt{G/F_{\rm n}}$ and $j_{\rm tot} \propto \left(G/F_{\rm n}\right)^2$.\\[12pt]

For the evaluation of damping and excitation due to evaporating particles, there is no such ambiguity over which case to take since we found that the excitation is 
the same for both the uniform $\theta=0$ rotation (old case) and isotropic $\theta$ distribution (new case).

\subsection{Excitation and damping: neutral grain, charged impactor}

We now consider the case of an ion impacting a neutral grain.  The analysis of evaporating particles is the same as that treated in Sections \ref{ss:nnd} and 
\ref{ss:nne}, 
since the ion is assumed to recombine on the grain surface and evaporate as a neutral.

Incoming ions follow a trajectory influenced by the dipole moment of the grain, both permanent and induced. The characteristic induced dipole energy for a grain with 
radius $R$ is $E_{\rm id}\sim q_{\rm e}^2R^{-1}$ (i.e. the attraction of the ion to the mirror charge).  For the PAH sequence, $R\approx 0.9N_{\rm C}^{1/2}\,$\AA, so 
we find $E_{\rm id}/k=1.7\times 10^5N_{\rm C}^{-1/2}\,$K.  Thus even at $N_{\rm C}\approx 100$ (our largest disc-like grains), the induced dipole energy is well 
above 
the temperature of the gas even in warm phases (WNM, WIM).  Therefore to a first approximation we treat the probability of an incoming ion striking the grain surface 
as being determined by the dipole interactions rather than grain geometry.  (Since the polarizability tensor of the grain is not isotropic, this is only an 
approximation.)  In this case, we are justified in using the AHD09 rates for incoming particles.  We are thus led to the conclusion that the AHD09 rates are 
applicable to ion impacts on neutral grains, both for the incoming coefficients $F,G_{\rm i}^{\rm(in)}$ and as previously described for the evaporation coefficients 
$F,G_{\rm i}^{\rm(ev)}$.


\subsection{Excitation and damping: charged grain, charged impactor}

In the case of an ion colliding with a charged grain, the particles interact with the Coulomb potential.
 which has magnitude 
\beq
V(r) = \frac{q_{\rm e}^2}{r}
\eeq
for single charges (and more for multiple charges).
A simple calculation then shows that for practical cases with the grains that are treated as disc-like ($a<6\,$\AA), ISM temperatures in most phases of interest 
(including warm phases) will have $\frac32kT\ll q_{\rm e}^2/a$.  In this case, the angular momentum transferred to the grain by incoming particles is 
geometry-independent: positive grains will receive essentially no impacts, while negative grains will be impacted by (and acquire the angular momentum of) any 
particle that passes close enough to the grain.  Thus the incoming rates $F,G_{\rm i}^{\rm(in)}$ are left unaffected.  The outgoing rates $F,G_{\rm i}^{\rm(ev)}$ are 
also unaffected: since the particles are neutral when they evaporate off the grain, the outgoing rates are as computed in the previous section.

\subsection{Summary}


We may now summarize the key differences between our investigation and that of AHD09.

For the case of neutral grains and neutral impactors, we have thus found the coefficients:
\beq
F_{\rm n} = \frac53F_{\rm n,AHD09} {\rm ~~and~~}
G_{\rm n} = G_{\rm n,AHD09}.
\eeq
For charged grains and neutral impactors,
\beq
F_{\rm n}^{\rm(in)} = \left\{1 + \frac23 \left[ 1 + \left( \frac{r_{\rm c}}{a_{\rm cx}} \right)^2 \right]^{-1} \right\}
F_{\rm n,AHD09}
\eeq
and
\beq
G_{\rm n} = G_{\rm n,AHD09}.
\eeq

The case of ion impacts is left unchanged from AHD09.

\section{Results}
\label{sec:results}

To avoid lengthy repetitions, we will refer to the case where grains are spinning around their axis of greatest inertia (as treated by DL98b and AHD09) by ``case 1'', and to the case where the relative 
orientation of the grain and the angular momentum is randomized (as dicussed in the present work) by ``case 2''.

\subsection{Angular momentum distribution}

We saw in Section~\ref{ss:em} that, at equal angular momentum, the total
power radiated by a disc-like grain in case 2 was
5 times (in the case $\mu_{\rm op}=0$) to $\sim 10$ times
($\mu_{\rm ip}^2:\mu_{\rm op}^2 = 2:1$) higher than the power radiated in case 1. This ratio
goes even higher as one increases the $\mu_{\rm op}^2:\mu_{\rm ip}^2$
ratio. However, the angular momentum distribution is different in each
case, and, as $P \propto L^4$, the ratio of the total power emitted will really be 
\beq
\frac{P_{\rm case 2}}{P_{\rm case 1}} \approx 10 \frac{\langle L^2\rangle^2_{\rm case 2}}{\langle L^2\rangle^2_{\rm case 1}}.
\eeq
In what follows we show that $\langle L^2\rangle_{\rm case 2} < \langle L^2\rangle_{\rm case 1}$.

\begin{figure} 
\begin{center}
\includegraphics[width = 3.2in]{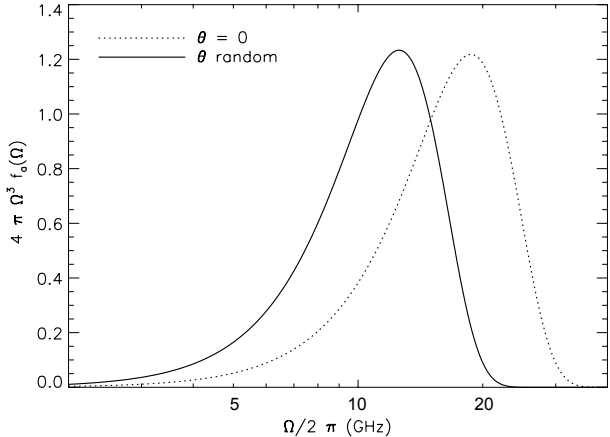}
\caption{Probability distribution function for the parameter $\Omega =
L/I_3$, for a grain of radius $a = 5\,$\AA, in WIM conditions, with
$\mu_{\rm ip}^2:\mu_{\rm op}^2 = 2:1$, and with dipole moment per atom
$\beta = 0.38$ debye. }
\label{fig:faOmega}
\end{center}
\end{figure}

First of all, we showed in earlier sections that the damping rates are
generally higher for grains spinning around a non principal axis. This
can be understood heuristically as follows: for a given angular
momentum $L$, the rotational energy $E_{\rm rot}(L,\theta)$ as a
function of the nutation angle was
given in Eq. (\ref{eq:Erot}). Averaging over angles, we find that
\beq
\langle E_{\rm rot} \rangle(L) = \frac{L^2}{2 I_1} -
\frac{1}{3}\frac{L^2}{2}\left(I_1^{-1} - I_3^{-1}\right).
\eeq
In the case of a disc-like grain,  ($I_3 = 2 I_1$) this is 
\beq
\langle E_{\rm rot} \rangle(L) = \frac53 \frac{L^2}{2 I_3} = \frac53 E_{\rm rot}(L,\theta=0).
\eeq
Therefore, we may expect that, when in contact with a bath of a
characteristic energy, grains with a randomly oriented rotation axis
will have an rms angular momentum $\sim \sqrt{5/3}$ times smaller
than those rotating around the axis of greatest
inertia. This is indeed what
we found in the case of collisions of neutral grains with neutral impactors, or emission of
infrared photons, for which we showed that $G$ was unchanged but $F$
was increased by a factor of $5/3$. We also showed that the normalized
plasma damping and excitation rates satisfied $1<F_{\rm p}/G_{\rm p}
<3$.

\begin{figure} 
\begin{center}
\includegraphics[width = 3.2in]{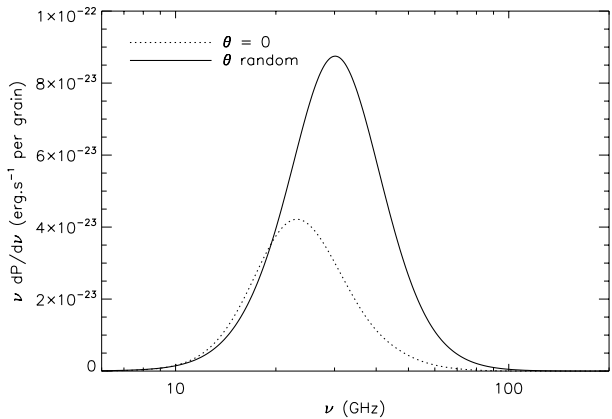}
\caption{Power radiated by a grain of radius $a = 5\,$\AA, in WIM conditions, with
$\mu_{\rm ip}^2:\mu_{\rm op}^2 = 2:1$, and with dipole moment per atom
$\beta = 0.38$ debye.}
\label{fig:power}
\end{center}
\end{figure}

\begin{figure} 
\begin{center}
\includegraphics[width = 3.2in]{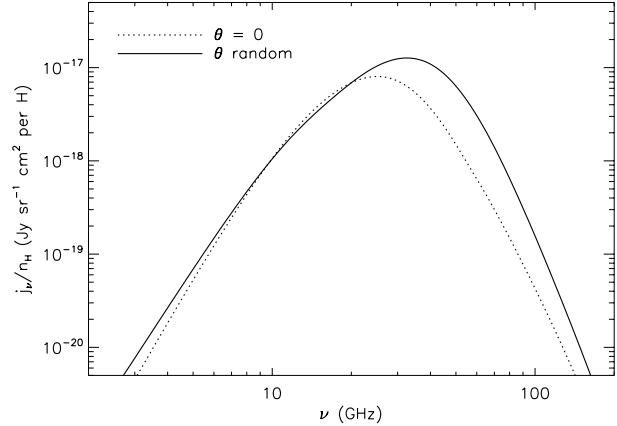}
\caption{Spinning dust emissivity in WIM environment.}
\label{fig:emissivity}
\end{center}
\end{figure}

More importantly, the characteristic radiation-reaction damping time
$\tau_{\rm ed}$ was
found to be shorter in case 2. We have
\beq
\frac{\tau_{\rm ed}(\theta \textrm{ random})}{\tau_{\rm ed}(\theta =
  0)} = \frac{\mu_{\rm ip}^2}{\frac{41}{15} \mu_{\rm ip}^2 +
  \frac{16}{3}\mu_{\rm op}^2}.
\eeq
In the case where radiation-reaction is the dominant rotational
damping mechanism, which is the case for the smallest grains in diffuse phases of the ISM, AHD09 showed that the rms angular momentum
is $\propto \tau_{\rm ed}^{1/4}$. Numerically, we have 
\beq
\frac{\tau^{1/4}_{\rm ed}(\theta \textrm{ random})}{\tau_{\rm ed}^{1/4}(\theta =
  0)}
\approx
\left\{\begin{array}{ll}
0.78 & \mu_{\rm op} = 0, \\
0.66 & \mu_{\rm ip}^2:\mu_{\rm op}^2 = 2:1. \end{array}\right.
\eeq
From these considerations, we therefore expect that in the same
environment, the characteristic angular momentum in case 2 will be $\sim$0.66--0.78 times the one in case 1.

We show in Fig.~\ref{fig:faOmega} the angular momentum distribution for a
grain of volume equivalent radius $a = 5$ \AA, in WIM conditions, with
$\mu_{\rm ip}^2:\mu_{\rm op}^2 = 2:1$, and with dipole moment per atom
$\beta = 0.38$ debye. The rms angular momentum in case 2 is $\sim 0.67$ times the one in case 1.

\subsection{Change in emissivity}

At a given angular momentum, the power radiated in case 2 peaks at a frequency approximately twice higher than the power radiated in case 1 (see discussion in Section 
\ref{ss:em}).

Therefore, and in view of the preceding section, we expect that the total power
radiated in case 2 will peak
at a frequency $\sim 2 \times 0.7 \sim 1.4$ times higher and will
integrate to a total power
$\sim 10 \times (0.7)^4 \sim 2$ times the power radiated in case 1. This is indeed what we find, as can be seen in Fig.~\ref{fig:power}.

\begin{figure*} 
\begin{center}
\includegraphics[width = 170mm]{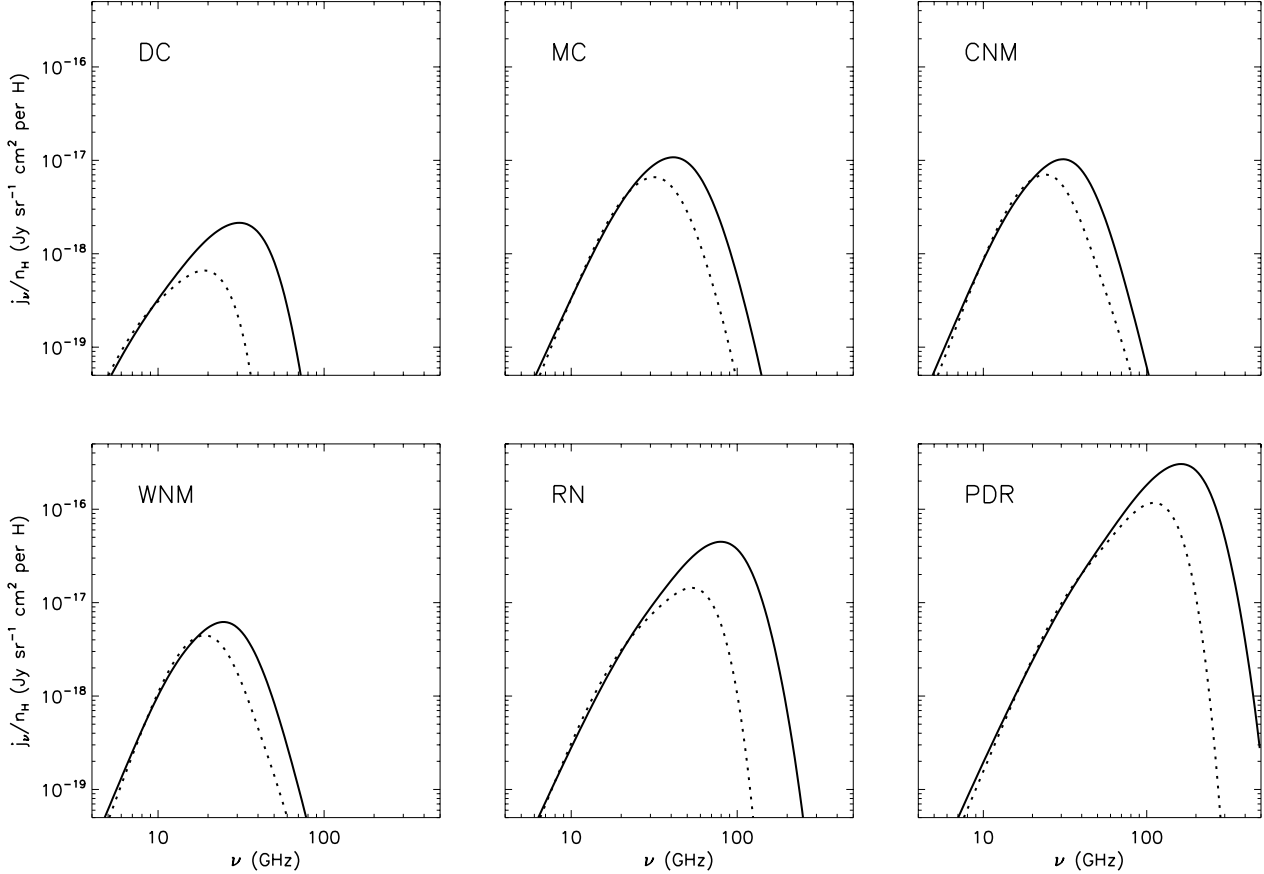}
\caption{Spinning dust spectra for several environmental conditions: dark cloud (DC), molecular cloud (MC), cold neutral medium (CNM), warm neutral medium (WNM), 
reflection nebula (RN) and photodissociation region (PDR). The environments are defined in DL98b, table 1. The parameters for the grain size distribution are $R_V = 
3.1, b_C = 6\times 10^{-5}$ for the diffuse CNM and WNM phases, and $R_V = 5.5, b_C = 3\times 10^{-5}$ for the dense DC, MC, RN and PDR. The dashed line is for a 
spectrum caculated assuming case 1 ($\theta=0$), whereas the solid line is for case 2 (isotropic $\theta$). The shift to higher frequencies and increase in emissivity 
in case 2 is systematic for all environments. We expect that case 2 should be a better approximation in the diffuse and high radiation intensity phases (WIM, CNM, 
WNM, RN, PDR).}
\label{fig:environments}
\end{center}
\end{figure*}

The overall spinning dust emissivity follows the same trends, as can be
seen in Fig.~\ref{fig:emissivity} for the WIM, and in Fig.~\ref{fig:environments} for other interstellar environments.

\subsection{Sensitivity to dipole moment orientation}

It is not clear what is the correct assignment for the direction of the grain permanent dipole moment relative to the principal axes.
Here we analyse the effect of the dipole moment orientation on the
spinning dust spectrum; it appears to make only a minor difference in the WIM environment.

\begin{figure} 
\begin{center}
\includegraphics[width = 3.2in]{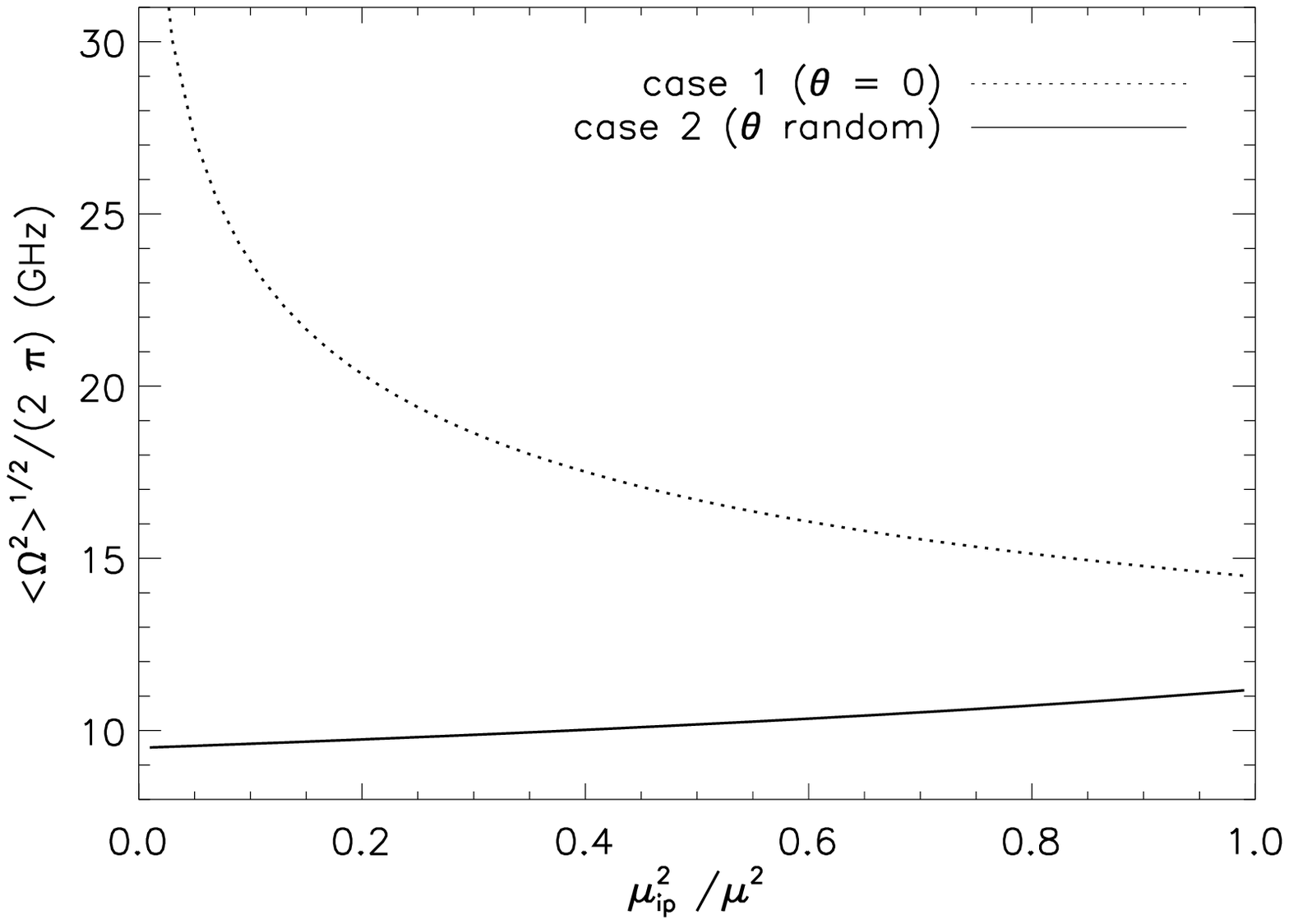}
\includegraphics[width = 3.2in]{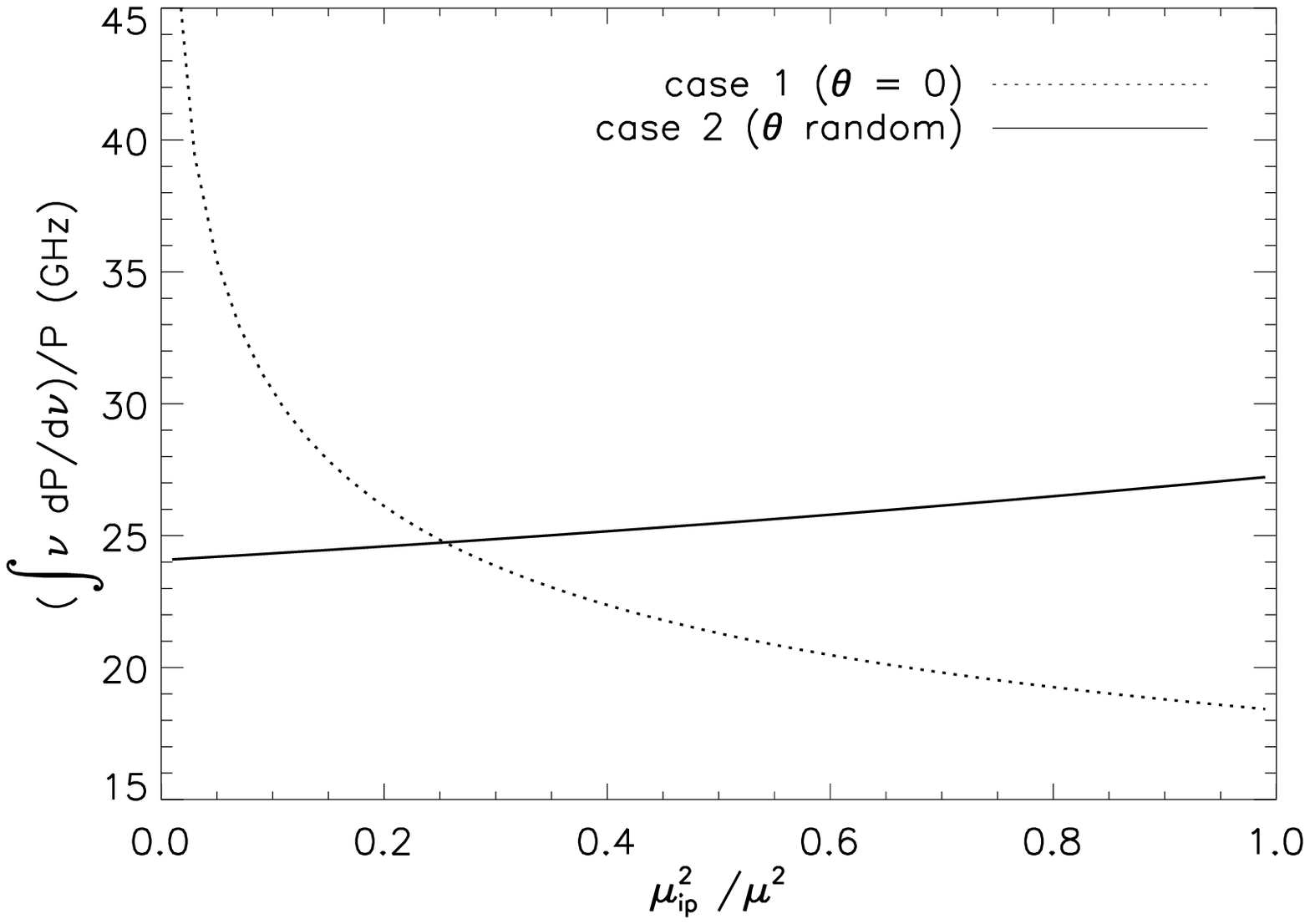}
\caption{Top panel: normalized rms angular momentum $\langle \Omega^2\rangle^{1/2}$ as a  function of the ratio of in-plane to total dipole moment. Bottom panel: 
estimate of the peak frequency $\int \nu (\rmd P/\rmd \nu) \rmd \nu \big{/} P$, as a function of this ratio. Both are for a dust grain of radius $a = 5\,$\AA\ and 
dipole moment per atom $\beta = 0.38$ debye, in WIM conditions.} \label{fig:rms_peak}
\end{center}
\end{figure}

For the smallest grains where radiation-reaction damping is most important, we expect $\langle \Omega^2\rangle^{1/2} \propto \tau_{\rm ed}^{1/4}$ so
\beq
\langle \Omega^2\rangle^{1/2} \propto
\left\{\begin{array}{ll}
\mu_{\rm ip}^{-1/2} & \textrm{(case 1)}, \\
\mu^{-1/2} \left(\frac{80}{39} - \frac{\mu_{\rm ip}^2}{\mu^2}\right)^{-1/4} & \textrm{(case 2)}.
\end{array}\right.
\eeq
In case 1 the rotation rate is very sensitive to the orientation of the dipole moment (only the in-plane component conributes to the power and the radiation reaction damping). Eventually, when the 
in-plane component becomes small enough, radiation-reaction damping becomes subdominant and the rms angular momentum will depend only on interactions with gas or infrared photons. In case 2 however the 
dependence on $\mu_{\rm ip}^2/\mu^2$ is quite weak, as the out-of-plane component contributes to the power and angular momentum loss. We show the normalized rms angular momenta in case 1 and 2 in
Fig.~\ref{fig:rms_peak}. Figure \ref{fig:rms_peak} also shows an estimate of the peak frequency of the emitted power in both cases.

\begin{figure}
\begin{center}
\includegraphics[width = 3.2in]{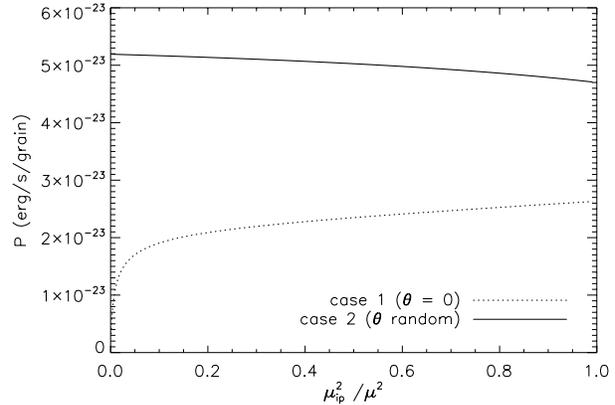}
\caption{Total power emitted by a dust grain of radius $a = 5\,$\AA\ and dipole moment per atom $\beta = 0.38$ debye, in WIM conditions,  as a  function of the ratio 
of in-plane to total dipole moment. }
\label{fig:pow_muip}
\end{center}
\end{figure}

The total power radiated by one grain, at a given angular momentum, was given in Eq.~(\ref{eq:totem}) for case 2. Taking $\Omega \sim \langle \Omega^2 \rangle^{1/2}$, and using the above results, we obtain
\beq
P \propto \left\{\begin{array}{ll}
{\rm constant} & \textrm{(case 1)} \\
\left(\frac{32}{17} - \frac{\mu_{\rm ip}^2}{\mu^2}\right) \Big{/}\left(\frac{80}{39} - \frac{\mu_{\rm ip}^2}{\mu^2}\right)
& \textrm{(case 2)}.
\end{array}\right.
\eeq
Thus in both cases the total power is very nearly independent of $\mu_{\rm ip}^2/\mu^2$. In case 1, when $\mu_{\rm ip}^2/\mu^2 \rightarrow 0$, radiation-reaction 
damping becomes subdominant and the power becomes proportional to $\mu_{\rm ip}^2$. These features are shown in Fig.~\ref{fig:pow_muip}.

\section{Discussion}
\label{sec:disc}

The purpose of this work was to revisit the assumption of DL98b and
AHD09 that PAHs rotate about their axis of main inertia. The
motivation in doing so is that thermal spikes following the absorption of
UV photons randomize the orientation of the grain with respect to the
angular momentum axis. These absorption events happen frequently enough (i.e. on timescale shorter than the timescale for significant changes in the total angular 
momentum) that we expect such a randomization to be effective in
most environments.  Thus we expect the results from this work (``case 2'') to be a better approximation to
diffuse or high-radiation environments (CNM, WNM, WIM, PDR, and RN) than those from AHD09, which assumed rapid dissipation of the nutational energy 
($\theta=0$ or ``case 1'').
However, the new release of {\sc SpDust} allows the user to choose either case; for example, one may
wish to explore the range of cases in dark cloud environments where thermal spikes are infrequent,
or what happens if an as-yet-unidentified dissipational process is active and restores $\theta=0$.


In this work, we showed that, for a given angular momentum, the power
radiated by a grain in case 2 is $\sim 10$ times higher than that
radiated by a grain in case 1. This is because in case 2, the grain
emits at higher frequencies, including above the one corresponding to the instantaneous angular velocity, as it is not rotating around the axis of
greatest inertia. 

We evaluated the rotational excitation and damping rates in case 2 as a function of grain size and environment conditions, and the resulting angular momentum 
distribution. We showed that in a given environment, grains in case 2 have a
lower rms angular momentum than those in case 1, by a factor of $\sim
0.7$. This is due to larger damping rates, in particular radiation-reaction damping, in case 2. 

The combination of these results leads to a spinning dust spectrum peaking at slightly
higher frequencies in case 2, and a total power approximately twice as large as
that emitted in case 1. Finally, we showed that the spectrum in case 2 is only weakly sensitive to the precise value of the $\mu_{\rm ip}^2:\mu_{\rm op}^2$ ratio.

\citet{2009ApJ...699.1374D} found a tension between theoretical results and microwave observations of the WIM: the theory was a factor of $\sim 3$ 
larger than the observations, and the peak frequency of the spinning dust and its amplitude could not be simultaneously reconciled by changing $\beta$ 
(the normalization of the dipole moment).  By increasing the theoretical emissivity and moving its peak to higher frequencies, our results may worsen 
this tension.  This seems likely to strengthen the empirical case for depletion of the PAH population in the WIM phase, however there are other 
conceivable explanations for this discrepancy.  The random walk model for the dipole moment may not apply well to the smallest grains (e.g. one could 
imagine that some of the small PAHs have symmetries that guarantee $\bmu=0$ exactly), or one could imagine
extra low-frequency internal degrees of freedom which allow the grain to relax to a state where it rotates around the axis of greatest moment of 
inertia.  A detailed 
exploration of the parameter space \citep[as was done by][]{2009ApJ...699.1374D} is beyond the scope of this paper.

As a final note, we present some of the remaining issues in the treatment of the rotational physics of the smallest dust grains:
\begin{itemize}
\item {\em Triaxiality}:
Many PAHs have triaxial moment of inertia tensors (e.g. ovalene C$_{32}$H$_{14}$, circumpyrene C$_{42}$H$_{16}$, and their derivatives).  This case was not treated in 
the present paper due to its much 
greater complexity: since the dipole moment then depends on elliptic functions of the angle conjugate to the nutation action, a countably infinite number of 
frequencies are emitted.  Aside from this aspect, however, the underlying formalism in this paper would be applicable: the nutation action (rather than $h K = 
2\pi L\cos\theta$) would be conserved in free rotation and we would average over this action instead of $\cos\theta$.  The analysis would also break into two cases 
depending on whether the grain lies on the short-axis or long-axis side of the separatrix.
\item {\em Impulsive torques}: Some of the sources of torque, such as ion impacts, impart large but infrequent changes in angular momentum.  This could in principle 
lead to ``rotational spikes'' analogous to the well-known thermal spikes in the grains' internal energy, and would not be treated correctly by the Fokker-Planck 
equation (which is a diffusive approximation).\footnote{This issue is treated in \citet{Hoang10}; they find that the principal effect on the spinning dust spectrum 
is the existence of a ``tail'' to high frequencies resulting from transient spin-up of the grains.}
\item {\em Ancillary data}: We have not fully quantified the uncertainties in the ancillary data, such as evaporation temperatures, the emissivity in the lowest-frequency vibrational modes, and the grain 
charging model (photoelectric and electron/ion impact).  However, our hope in making the {\sc SpDust} code publicly available is to provide users the flexibility to explore 
deviations from default or fiducial parameters.
\end{itemize}

\section*{Acknowledgments}

We thank Bruce Draine and Nathalie Ysard for numerous conversations about the physics of grain rotation, and
Clive Dickinson for reading the manuscript and testing {\sc SpDust} v2.0.

K. S. would like to thank Edward C. and Alice Stone for their support for the Summer Undergraduate Research Fellowship programme at Caltech.
Y. A-H. and C. H. are supported by the U.S. Department of Energy (DE-FG03-92-ER40701) and the National Science Foundation (AST-0807337).
C. H. is supported by the Alfred P. Sloan Foundation.

\appendix

\section{Implementation of plasma drag integrals}
\label{app:Gp}

Here we describe our implementation of the plasma drag coefficients, Eqs.~(\ref{eq:Gp}) and (\ref{eq:Fp}), in {\sc SpDust}.  These are integrals over the $G_{\rm p, 
\rm AHD}(\omega)$ function, which is itself time-consuming to compute.

If we wish to calculate the integral $\int w(x) g(x) \rmd x$,
where $w(x)$ is a known weighing function which propeties will be discussed later, and the function $g(x)$ is smooth enough on the interval of integration that it can 
be approximated by a quadratic polynomial
$g(x) \approx a + b x + c x^2$,
then we may approximate
\beq
\int w(x) g(x) \rmd x \approx A\left[ g(x_+) + g(x_-)\right],
\label{eq:wg-approx}
\eeq
where $A \equiv \frac{1}{2} \int w(x) \rmd x$, and $\{x_+, x_-\}$ are the solutions of the second order system 
\beq
\left\{\begin{array}{rcl}
x_+ + x_- &=& A^{-1} \int x w(x) \rmd x, \\
x_+^2 + x_-^2 &=& A^{-1} \int x^2 w(x) \rmd x .\end{array}\right.
\eeq


We now turn our attention to the specific cases of $G_{\rm p}(\Omega)$ and $F_{\rm p}(\Omega)$.
With $x = \omega/\Omega$ and the weighing function $w(x) = (3 - x)^2$, we get
\barr
G_{\rm p}(\Omega) &\approx& \frac{2 \mu_{\rm op}^2}{3 \mu_{\rm ip}^2}  G_{\rm p, \rm AHD}(2 \Omega)\nonumber \\
&& + \frac{1}{3}\left[G_{\rm p, \rm AHD}(\Omega_+) +G_{\rm p, \rm AHD}(\Omega_-)\right],
\earr
where 
\beq
\Omega_{\pm} =  \frac{3 \pm \sqrt{3/5}}{2} \Omega \approx \{1.11 \Omega, 1.89 \Omega\}.
\eeq

Similarly, with the weighing function $ w(x) = x(3 - x)^2$, we get
\barr
F_{\rm p}(\Omega) &\approx& \frac{4 \mu_{\rm op}^2}{3 \mu_{\rm ip}^2}  G_{\rm p, \rm AHD}(2 \Omega)\\
&& + \frac{1}{2}\left[G_{\rm p, \rm AHD}(\tilde \Omega_+) +G_{\rm p, \rm AHD}(\tilde\Omega_-)\right],
\earr
where
\beq
\tilde\Omega_{\pm} = \frac{8 \pm \sqrt{13/3}}{5} \Omega \approx \{ 1.18 \Omega, 2.02 \Omega\}.
\eeq

We have tested the accuracy of the approximate integrator and found that the error was less than 1\% in the regime where $F_{\rm p}, G_{\rm p}$ have significant 
values, 
i.e. for $\Omega \lesssim \Omega_{\rm th} = \sqrt{3 k T/I_3}$. More precisely, we checked that
\beq
\frac{|\Delta F_{\rm p}(\Omega)|}{F_{\rm p}(\Omega)} \times \min\left(1, \frac{F_{\rm p}(\Omega)}{F_{\rm p}(\Omega_{\rm th})}\right) < 0.01
\eeq
for grain radii $a = 4, 5, 6$ \AA, gas temperatures $T = 50, 500, 5000$ K and grain charge $Z = -1, 0, 1$, and similraly for $G_{\rm p}$.

\label{lastpage}

\end{document}